\documentclass[a4paper,fleqn,usenatbib]{mnras}
\usepackage[english]{babel}
\usepackage[utf8x]{inputenc} 
\usepackage{newtxtext,newtxmath}
\usepackage{longtable}
\usepackage{setspace}
\usepackage{booktabs}
\usepackage{wallpaper}
\usepackage{epstopdf}
\usepackage{eso-pic}

\usepackage[singlelinecheck=off]{caption}
\usepackage{longtable}

\usepackage[T1]{fontenc}
\usepackage{ae,aecompl}

\usepackage{graphicx}	
\usepackage{amsmath}	
\usepackage{amssymb}	

\def\HII{H{\sc ii}}

\title[N-Fractionation Gradient]{Nitrogen fractionation in high-mass star-forming cores across the Galaxy}

\author[L. Colzi et al.]{L. Colzi$^{1,2}$\thanks{E-mail: colzi@arcetri.astro.it},
F. Fontani$^{2}$,
V. M. Rivilla$^{2}$,
A. Sánchez-Monge$^{3}$,
L. Testi$^{2, 4}$,
\newauthor M. T. Beltrán$^{2}$,
and P. Caselli$^{5}$
\\
$^{1}$Università degli studi di Firenze, Dipartimento di Fisica e Astronomia, Via Sansone 1, 50019 Sesto Fiorentino, Italy\\
$^{2}$INAF-Osservatorio Astrofisico di Arcetri, Largo E. Fermi 5, I-50125, Florence, Italy    \\
$^{3}$I. Physikalisches Institut of the Universität zu Köln, D-50937 – Cologne, Germany\\
$^{4}$ESO, Karl Schwarzschild str. 2, D-85748 Garching, Germany\\
$^{5}$Max-Planck-Instit\"{u}t f\"{u}r extraterrestrische Physik, Giessenbachstrasse 1, D-85748, Garching bei M\"{u}nchen, Germany 
}

\date{Accepted 2018 April 13. Received 2018 April 13; in original form 2017 December 24}

\pubyear{2017}

\begin{document}
\label{firstpage}
\pagerange{\pageref{firstpage}--\pageref{lastpage}}
\maketitle

\begin{abstract}
The fractionation of nitrogen (N) in star-forming regions is a poorly understood process.
To put more stringent observational constraints on the N-fractionation, we have observed with the IRAM-30m telescope a large sample of 66 cores in massive star-forming regions. We targeted the (1--0) rotational transition of HN$^{13}$C, HC$^{15}$N, H$^{13}$CN and HC$^{15}$N, and derived the $^{14}$N/$^{15}$N ratio for both HCN and HNC. We have completed this sample with that already observed by Colzi et al.~(\citeyear{colzi18}), and thus analysed a total sample of 87 sources. 
The $^{14}$N/$^{15}$N ratios are distributed around the Proto-Solar Nebula value with a lower limit near the terrestrial atmosphere value ($\sim$272). 
We have also derived the $^{14}$N/$^{15}$N ratio as a function of the Galactocentric distance and deduced a linear trend based on unprecedented statistics. The Galactocentric dependences that we have found are consistent, in the slope, with past works but we have found a new local $^{14}$N/$^{15}$N value of $\sim$400, i.e. closer to the Prosolar Nebula value. A second analysis was done, and a parabolic Galactocentric trend was found. Comparison with Galactic chemical evolution models shows that the slope until 8 kpc is consistent with the linear analysis, while the flattening trend above 8 kpc is well reproduced by the parabolic analysis. 
\end{abstract}



\begin{table}
\begin{center}
\caption{Line rest frequencies and observational parameters. {\it F} is the quantum number associated with the sum between the total angular momentum {\it |J|} and the $^{14}$N nuclear angular momentum.}
  \begin{tabular}{lcccc}
  \hline
  Line   & Hyperfine& Frequency & log$_{10}$({\it A}$_{\rm ij}$*) & {\it E}$_{\rm U}$**\\ 
          & Component &    (GHz)              &                  &        (K)     \\
  \hline
H$^{13}$CN(1--0) & F= 1--1 &86.3387$^{1}$  &    -4.62444***     &   4.1 \\
 & F= 2--1 & 86.3402$^{1}$  &         &   \\
 & F=0--1 &86.3423$^{1}$  &        &    \\
 HN$^{13}$C(1--0) & & 87.0908$^{2}$  &       -4.72878  &  4.2 \\
HC$^{15}$N(1--0)  & & 86.0549$^{3}$ &     -4.62943    &  4.1\\
H$^{15}$NC(1--0)  & & 88.8657$^{4}$ &       -4.70235  & 4.3  \\
  \hline
  \normalsize
  \label{observation}
  \end{tabular}
  \end{center}
  \begin{flushleft}
 \footnotesize
 *Einstein coefficient of the transition;\\
 **Energy of the upper level;\\
 ***It refers to the whole transition;\\
   $^{1}$ Cazzoli \& Puzzarini (\citeyear{cazzoli05b});\\
    $^{2}$ van der Tak et al.~(\citeyear{tak09});\\
     $^{3}$ Cazzoli et al.~(\citeyear{cazzoli05a});\\
      $^{4}$ Pearson et al.~(\citeyear{pearson76}).
        \end{flushleft}
       \normalsize
\end{table}

 \section{Introduction}
 \label{intro}
 Nitrogen, the fifth most abundant element in the Universe and the fourth most important biogenic element, exists in the form of two stable isotopes: $^{14}$N (the main one) and $^{15}$N. The $^{14}$N/$^{15}$N ratio is thought to be an important indicator of the chemical evolution of the Galaxy, although its value across the Galaxy is still uncertain. The ratio measured for the Proto-Solar Nebula (PSN), from the Solar wind, is 441$\pm$6 (Marty et al.~\citeyear{marty10}), and this value is about two times larger than that measured in the terrestrial atmosphere (TA), derived from N$_{2}$, $\sim$272 (Marty et al.~\citeyear{marty09}). The PSN value is also larger than that measured in some comets (a factor two) and in carbonaceous chondrites (until a factor five). A $^{14}$N/$^{15}$N ratio of 139$\pm$26, from HCN, was estimated in the comet 17P/Holmes after an outburst (Bockelée-Morvan et al.~\citeyear{bockelee-morvan08}). Manfroid et al.~(\citeyear{manfroid09}) estimated a mean value of $^{14}$N/$^{15}$N=148$\pm$6 from CN in 18 comets observed with VLT+UVES. Different values of $^{14}$N/$^{15}$N were estimated in the carbonaceous chondrite Isheyevo: $\sim$424 in a osbornite-bearing calcium-aluminium-rich inclusion (CAI) (Meibom et al.~\citeyear{meibom07}), and from 44 up to 264 (van Kooten et al.~\citeyear{vankooten17}) in lethic clasts.
Measurements in carbonaceous chondrites are made in specific parts of the objects, as described above, while measurements in comets refers to values integrated over a whole part of comets's comas. For these reasons values obtained from extraterrestrial material studied on Earth could be different from observations of comets. At the same time, different values from different part of the same chondrite are measured, and this indicated that different part of a chondrite are formed during different time of the star formation. Meibom et al.~(\citeyear{meibom07}) infer that the carbon-bearing titanium-nitride (osbornite) in the Isheyevo CAI formed by gas-solid condensation, started in a high-temperature ($\sim$2000 K) inner region (<0.1 AU) of the PSN, where all solids were initially evaporated and the gas homogenized. Under such high temperature conditions, little isotopic fractionation is expected between gas and solids, and the nitrogen isotopic composition of osbornite in the Isheyevo CAI must be representative of the Solar nebula. In fact, winds associated with bipolar outflows, or turbulent transport of the hot inner nebula (silicate) dust, may have carried small refractory particles out to colder zones where CAI formed during protoplanet formation. van Kooten et al.~(\citeyear{vankooten17}) infer, from isotope data of Isheyevo lethic clasts, that variations in N isotopes are consistent with the accretion of multiple organic precursors and subsequent alteration by fluids with different (isotopic) compositions, that is in stages more evolved than the enrichments in CAI.
 
The relation between the $^{15}$N-enrichments in pristine Solar System materials and the natal core is still uncertain. The number of observational works is increasing, but observations towards both low- and high-mass star-forming cores can not reproduce the low $^{14}$N/$^{15}$N values found in comets or meteorites (e.g. Fontani et al.~\citeyear{fontani15}, Colzi et al.~\citeyear{colzi18}, Zeng et al.~\citeyear{zeng17}, Daniel et al.~\citeyear{daniel16}, Kahane et al.~\citeyear{kahane18}). Recent observations of PSN analogs, demonstrated that multiple isotopic reservoirs were present in the early phases of the formation of the Solar System (Hily-Blant et al.~\citeyear{hily-blant17}). In low-mass pre-stellar cores or protostellar objects, values of $^{14}$N/$^{15}$N were found comparable with that of the PSN: 330$\pm$150 from N$_{2}$H$^{+}$ (Daniel et al.~\citeyear{daniel16}), 350 -- 850 from NH$_{3}$ (Gerin et al.~\citeyear{gerin09}), 334$\pm$50 (Lis et al.~\citeyear{lis10}), and an average value of 270$\pm$20 from different molecules, CN, HCN, HNC, HC$_{3}$N and N$_{2}$H$^{+}$ (Kahane et al.,~\citeyear{kahane18}) or even larger, 1000$\pm$200, in N$_{2}$H$^{+}$ (Bizzocchi et al.~\citeyear{bizzocchi13}). On the contrary, observations of nitrile-bearing species such as HCN and HNC in low-mass sources show lower values of the $^{14}$N/$^{15}$N ratio (140--360, Hily-Blant et al.~\citeyear{hily-blant13}; 160--290, Wampfler et al.~\citeyear{wampfler14}), but still not as low as in comets or protoplanetary discs ( 80--160, Guzmán et al.~\citeyear{guzman17}). Daniel et al.~(\citeyear{daniel13}) measured the nitrogen isotopic ratio in several nitrogenated species towards B1b. They found 300$^{+55}_{-40}$ from NH$_{3}$, 400$^{+100}_{-65}$ from N$_{2}$H$^{+}$, 330$^{+60}_{-50}$ from HCN, 225$^{+75}_{-45}$ from HNC and 290$^{+160}_{-80}$ from CN.

The $^{15}$N-enrichment is still not well understood, even from a theoretical point of view. Wirström et al.~(\citeyear{wirstrom12}) suggested that the isotopic enrichments measured in the primitive organic matter probably had their chemical origin in a low-temperature environment in which ion-molecule isotopic exchange reactions are favoured. Roueff et al.~(\citeyear{roueff15}) discusses some possible chemical pathways in cold ({\it T}$=$10 K) and dense gas ({\it n({\rm H}$_{2}$)}=2$\times$10$^{4-5}$cm$^{-3}$) that bring $^{15}$N-enrichments in molecules such as N$_{2}$H$^{+}$, HCN and HNC. However, the $^{14}$N/$^{15}$N found in the model after 10$^{7}$ yr is not different from that of the PSN value, and chemical reactions that bring $^{15}$N in molecules are uncertain or incomplete. Therefore, it seems that time does not play a role in N-fractionation, as found by Colzi et al.~(\citeyear{colzi18}) in a sample of massive cores in different evolutionary stages.
In fact, comparing the $^{14}$N/$^{15}$N ratios derived from HCN and HNC for these different high-mass star-forming cores, they do not find any trend with the evolutionary stages. This result suggests that chemical reactions may still be missing in existing models, or that the enrichment is a local phenomenon occurring in a spatial region much smaller than the beam size of the observations.

The abundance ratio $^{14}$N/$^{15}$N is also considered a good indicator of stellar nucleosynthesis since the two elemental isotopes are not originated in the same way. Both isotopes have indeed an important secondary production in the CNO cycles. There are two types of CNO cycles: a cold cycle and a hot cycle. The cold CNO cycle takes place in main-sequence stars and in the H-burning shells of red giants: $^{14}$N is created from $^{13}$C or $^{17}$O and brought in stellar surfaces through dredge-up on the red giant branch. The hot CNO cycle occurs instead in novae outbursts (Clayton~\citeyear{clayton03}), and is the main way to produce $^{15}$N.
However, there is also a strong primary component of $^{14}$N created in the so-called Hot Bottom Burning (HBB) of asymptotic giant branch (AGB) stars (e.g. Schmitt \& Ness~\citeyear{schmitt02}, Izzard et al.~\citeyear{izzard04}), and an (over-)production of $^{15}$N with respect to $^{14}$N in the relative role played by massive stars and novae (Romano \& Matteucci \citeyear{romano03}, Romano et al.~\citeyear{romano17}). Therefore, $^{15}$N is principally a secondary element.

Different origins of the two nitrogen isotopes lead to an increase of $^{14}$N/$^{15}$N ratio with the Galactocentric distance, up to a distance of $\sim$8 kpc, as predicted by models of Galactic chemical evolution (GCE, Romano \& Matteucci \citeyear{romano03}, Romano et al.~\citeyear{romano17}). However, the relative importance of these processes, or the eventual existence of additional processes, are still unclear. The only way to test this is to provide more observational contraints.
The GCE model made by Romano et al.~(\citeyear{romano17}) was compared  with measurements of the $^{14}$N/$^{15}$N ratios in a sample of warm molecular clouds observed by Adande \& Ziurys (\citeyear{adande12}). The model seems able to reproduce the trend found by Adande \& Ziurys (\citeyear{adande12}), but this latter was obtained from a small sample (22 sources), and mixing up data from different molecules and different instruments (i.e. different observational parameters).

In this work we report the measurements of the $^{14}$N/$^{15}$N ratio derived in a sample of 66 dense cores that span Galactocentric distances in the range $\sim$ 2--12 kpc. Moreover, together with the sources of Colzi et al.~(\citeyear{colzi18}), we present a new Galactocentric behavior of $^{14}$N/$^{15}$N, based now on 87 sources, which makes this new Galactocentric dependence statistically more robust than that of past works. We compare also this trend to GCE models. The results are presented in Sect.~\ref{res} and a discussion of the results is presented in Sect.~\ref{disc}.
\begin{figure*}
\centering
\includegraphics[width=30pc]{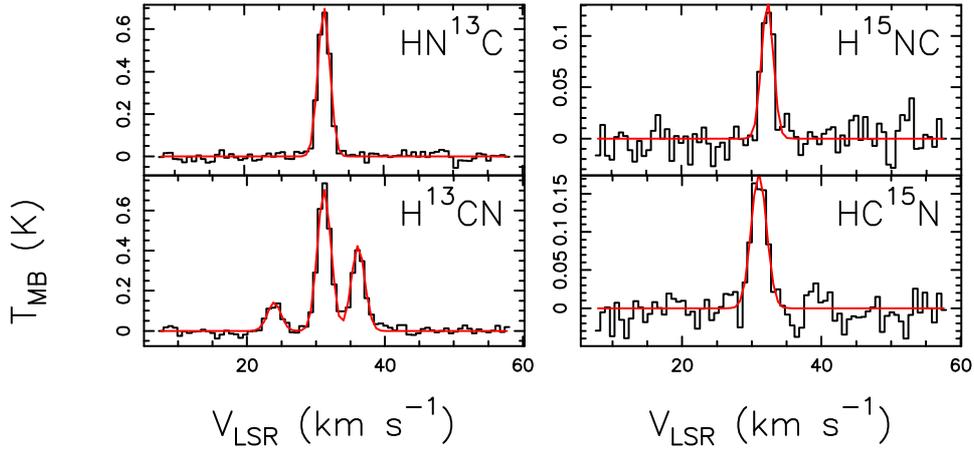}
\caption{Spectra of the source 18151-1208M1 in all the transitions detected: HN$^{13}$C(1--0) (top-left), H$^{15}$NC(1--0) (top-right), H$^{13}$CN(1--0) (bottom-left) and HC$^{15}$N(1--0) (bottom-right). For each spectrum the x-axis represents the systemic velocity listed in Tables \ref{fit1} and \ref{fit2}. The y-axis shows the intensity in main beam temperature units. The red curves are the best fits obtained with MADCUBA (from the method described in Sect.~\ref{res}).}
\centering
\label{text-spectra}
\end{figure*}

 \section{Source selection and observations}
 \label{observations}
 \subsection{Source sample}
The source list of the 66 sources is reported in Table \ref{tab-coo}, where the source coordinates and the distances from the Sun are shown. 
A part of the source sample was compiled from a list of high-mass young stellar objects from northern millimeter continuum surveys (Sridharan et al.~\citeyear{sridharan02}, Beuther et al.~\citeyear{beuther02}). We selected 35 sources with (i) distances < 5 kpc, to sample comparable spatial scales, (ii) strong (peak intensity > 0.5 Jy/beam) and compact 1.2 mm continuum emission, suggestive of high-mass young objects embedded in dust, and (iii) observable with the IRAM-30m telescope. Note that we do not consider any criteria related to the line intensity or molecular richness of the source, nor do we consider the centimeter continuum emission and luminosity, in order to avoid possible biases. The sources cover a broad range of luminosities and evolutionary stages (traced with different star formation signposts, e.g. presence of masers, infrared sources, HII regions) which will permit to study the dependence of chemical properties with luminosity and evolutionary stage. This sample is not classified in evolutionary stages (like that of Colzi et al.~\citeyear{colzi18}), yet. This classifation will be done in a forthcoming paper (Mininni et al., in prep). The whole sample and their properties will be presented in the forthcoming paper, while in this paper we focus on the lines of the isotopologues of HCN and HNC. The other 31 sources, indicated with $^{**}$ in Table \ref{tab-coo}, were selected from the literature (Fontani et al.~\citeyear{fontani11}, Wouterloot et al.~\citeyear{wouterloot93}, Beltrán et al.~\citeyear{beltran16}, Cesaroni et al.~\citeyear{cesaroni17}, Tan et al.~\citeyear{tan14}).

The distances are kinematic distances based on the rotation curve of the Galaxy, and are taken from Sridharan et al.~(\citeyear{sridharan02}), Beuther et al.~(\citeyear{beuther02}) and from the other works cited above.
For the sources with the ambiguity of distance we have chosen the near one. 

The sample of 66 sources was implemented with that observed by Colzi et al.~(\citeyear{colzi18}) of 27 sources, to obtain a statistically significant sample.   The sample of 27 sources was divided in the three evolutionary stages: 11 high-mass starless cores (HMSCs), 9 high-mass protostellar objects (HMPOs), and 7 ultra-compact \HII\ regions (UC HIIs), respectively (see Fontani et al.~\citeyear{fontani11}). Because 6 sources of the two samples were in common: 00117+6412M1, 18089-1732M1, 18517+0437M1, G75, 20081+3122M2 and NGC7538IRS9, we have decided to keep the results found in Colzi et al.~(\citeyear{colzi18}) as the S/N of their observations is better than that of the observations presented here by a factor of $\sim$2 (compare spectra in Appendix \ref{ap-spectra} of this paper and Appendix B of Colzi et al.~\citeyear{colzi18}). 
Despite the ratios found in Colzi et al.~(\citeyear{colzi18}) were derived using the kinetic temperatures calculated from ammonia (given in Fontani et al.~\citeyear{fontani11}), that are different from the excitation temperature ({\it T}$_{\rm ex}$) adopted in this work, the $^{14}$N/$^{15}$N ratios are almost independent (differences of 1--10\%) from the chosen temperature (see Sect.~\ref{res}), and then their results could be used to increment our sample.

We finally have a sample of 87 sources: the 27 of Colzi et al.~(\citeyear{colzi18}) and the 60 of this work not in common with the previous one.

 \subsection{Observations}
The 66 sources analyzed in this work were observed with the IRAM 30m telescope on August 2014. We have used the FTS spectrometer (Fast Fourier Transform Spectrometer; Klein et al.~\citeyear{klein12}) to cover the frequency range 85.7--93.5 GHz with a frequency resolution of 200 kHz, corresponding to 0.6 km s$^{-1}$ at the observed frequencies. The half power beam width (HPBW) is about 29 arcsec. The observations were made in position-switching mode. Pointing was checked every 1.5 hours, and focus was corrected at the beginning of the observations and every 4-6 hours. The integration time ranged from 10 to 30 minutes, except for some sources, as 04579+4703, for which it was less (about 5 minutes). The r.m.s. noise levels of the spectra are presented in Appendix \ref{fit1}, \ref{fit2}, \ref{fit3} and \ref{fit4}. The system temperature was in the range 100--150 K during the observations. The observed spectra were first calibrated and reduced using the CLASS/GILDAS package (Pety et al.~\citeyear{pety05}), the data were corrected for platforming and spike channels were removed. 
The beam efficiency ($B_{\rm eff}$) for the IRAM 30m telescope at these frequencies is 0.86, while the forward efficiency ($F_{\rm eff}$) is 0.95. The antenna temperatures were converted to main beam temperatures by using the expression: $T_{\rm A}^{*}$ = $T_{\rm MB} \eta _{\rm MB}$, where $\eta_{\rm MB}$ = $B_{\rm eff}$/$F_{\rm eff}$.
 The spectra obtained were exported from the software package CLASS of GILDAS\footnote{The GILDAS
software is available at http://www.iram.fr/IRAMFR/GILDAS} to MADCUBA\footnote{Madrid Data Cube Analysis on ImageJ is a software developed in the
Center of Astrobiology (Madrid, INTA-CSIC) to visualize and analyze
single spectra and datacubes (Martín et al., in prep.).} (see e.g.  Rivilla et al.~\citeyear{rivilla16},  Rivilla et al.~\citeyear{rivilla17}). With MADCUBA we have selected the part of the spectra around the four targeted lines, and we have fitted the baselines with a maximum order of 1. MADCUBA was then used to identify the lines and analyze them. For the analysis we have used the spectroscopic parameters from the Jet Propulsion Laboratory (JPL) molecular catalog\footnote{http://spec.jpl.nasa.gov/}. Rest frequencies are taken from the laboratory works and quantum-chemical calculations cited in Table \ref{observation}. 


\begin{figure*}
\centering
\includegraphics[width=40pc]{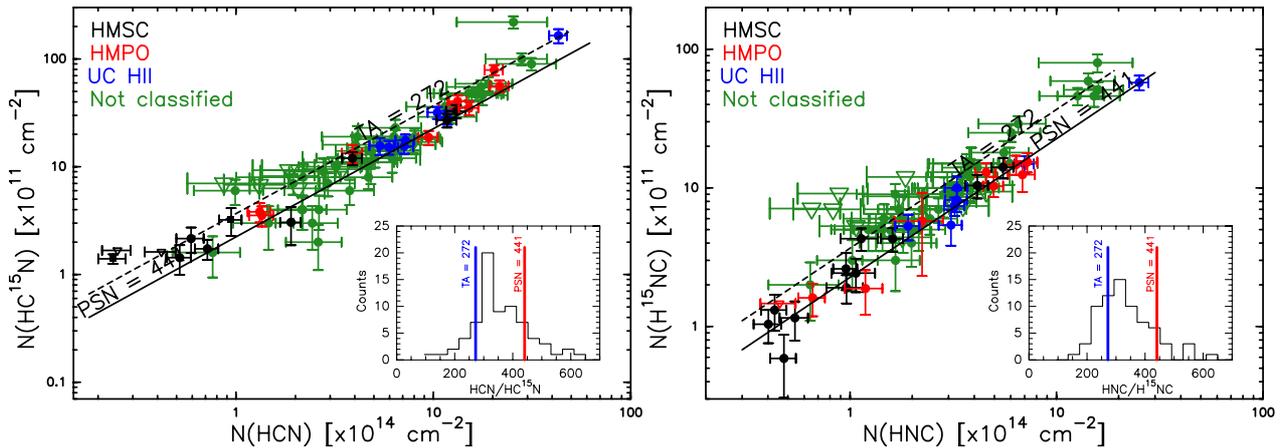}
\caption{Column densities of HCN against those of HC$^{15}$N (left panel) and of HNC against those of H$^{15}$NC (right panel). HCN and HNC column densities are derived from those of H$^{13}$CN and HN$^{13}$C, respectively (see Sect.~\ref{res}). In all panels the filled circles represent the detected sources and the open triangles are the upper limits for the corresponding column densities. In green the 60 sources of this work, in black, red and blue the 27 sources of Colzi et al.~(\citeyear{colzi18}) divided in the three evolutionary stages: high-mass starless cores (HMSCs), high-mass protostellar objects (HMPOs), and ultra-compact \HII\ regions (UC HIIs), respectively. The dashed line indicates the mean atomic composition as measured in the terrestrial atmosphere (TA) and the solid line that measured in the Proto-Solar Nebula (PSN). In the lower right corner of each panel we show the distribution of $^{14}$N/$^{15}$N values for HCN (left) and HNC (right). Here the blue vertical line represents the TA value and the red vertical line the PSN value.}
\centering
\label{fig-colden}
\end{figure*}

  \section{Analysis and results}
 \label{res}
 
 \subsection{Line detection and fitting procedure}
 \label{fitting}
  We have considered detections of HN$^{13}$C(1--0), H$^{15}$NC(1--0), H$^{13}$CN(1--0), and HC$^{15}$N(1--0) if the spectral lines show a peak main beam temperature ({\it T}$_{\rm MB}^{\rm peak}$)$\ge$3$\sigma$, where $\sigma$ is the r.m.s noise of the spectrum; for the detections close to the 3$\sigma$ limit, we have distinguished between tentative detections ($2.5\sigma \leq T_{\rm MB}^{\rm peak}< 3\sigma$) and non detections ($T_{\rm MB}^{\rm peak}<2.5\sigma$). For the non detections we have computed the upper limits for the column densities as explained in Colzi et al.~(\citeyear{colzi18}), with average value of the full width at half maximum ($\Delta v_{\rm 1/2}$) of the lines: $\Delta v_{{\rm HN}^{13}{\rm C}}=2.6\pm 0.1$ km s$^{-1}$, $\Delta v_{{\rm H}^{13}{\rm CN}}=3.3\pm 0.2$ km s$^{-1}$, $\Delta v_{{\rm H}^{15}{\rm NC}}=2.3\pm 0.1$ km s$^{-1}$ and $\Delta v_{{\rm HC}^{15}{\rm N}}=3.1\pm 0.2$ km s$^{-1}$.
 The percentage of detected lines (together with tentative tetections) is the following: the HN$^{13}$C line is clearly detected in 58 cores (88\%), the H$^{13}$CN line is detected in 58 cores (88\%) which, however, are not exactly the same (see Tables \ref{fit1} and \ref{fit2}), the H$^{15}$NC line is detected in 45 cores (68\%) and the HC$^{15}$N line is detected in 49 cores (74\%). 
 
Both HN$^{13}$C(1--0) and H$^{13}$CN(1--0) have hyperfine structure, 
but we can resolve it only for H$^{13}$CN(1--0) because the line widths found ($\sim$3 km s$^{-1}$) are smaller than the separations in velocity of the hyperfine components (e.g. Fig.~\ref{text-spectra}). Conversely, H$^{15}$NC(1--0) and HC$^{15}$N(1--0) do not have hyperfine structure. All the lines were fitted with single Gaussians, except in the case of H$^{13}$CN(1--0) for which we take into account the hyperfine structure, using same {\it T}$_{\rm ex}$ and $\Delta v_{1/2}$ for each transition. Fitting the HN$^{13}$C(1--0) with a single Gaussian can overestimate the line width. To estimates how much the line widths are overestimated by this simplified approach, when possible, we fitted a line both with one gaussian, and considering the hyperfine structure. We found that with a single Gaussian, the line width is about 10\% larger than with the hyperfine structure method. 

One of the sources, G31.41+0.31, requires some comments: it is a prototypical hot molecular core (HMC) which harbours deeply embedded young stellar objects (YSOs), and the HMC is separated by $\sim$5'' from an ultracompact \HII\ region (UC HII). Observations with the IRAM Plateau de Bure demonstrated the existence of a velocity gradient across the core (Beltrán et al.~\citeyear{beltran04} and \citeyear{beltran05}). Subsequently, observations with the Submillimeter Array confirmed the velocity gradient (Cesaroni et al.~\citeyear{cesaroni11}), supporting the claim of the presence of a rotating toroid. Moreover, observations of $^{12}$CO and $^{13}$CO revealed self-absorption caused by infalling gas close to the HMC (Cesaroni et al.~\citeyear{cesaroni11}). Therefore, G31.41+0.31 is a very complicated source and, as can be noted from the spectra in Appendix \ref{ap-spectra}, there is an evidence of possible self-absorption at the center of the lines (for H$^{13}$CN and HN$^{13}$C in particular) and a distorted line profile due to the possible presence of the rotating toroid at larger scales, together with the possibility of outflows from the center of the HMC (Cesaroni et al.~\citeyear{cesaroni17}, Beltrán et al.,~\citeyear{beltran18}). 
For these reasons we were not able to fit the lines neither with a single Gaussian nor with the three hyperfine components, because of the complex line shape. All of this makes it very difficult to fit the lines with the approximation used for the other sources; therefore, we decided to exclude this source from the analysis.

For G14.99-0.67, instead, we have identified two velocity components, and fit them simultaneously (G14.99-0.67 and G14.99-2). 

All the spectra are shown in Appendix \ref{ap-spectra}.

\subsection{Column density calculation}
To compute the colum densities we have used MADCUBA assuming local thermal equilibrium (LTE) conditions. MADCUBA takes five parameters to model the LTE line profile into account: total column density of the molecule ({\it N}), {\it T}$_{\rm ex}$, $\Delta v_{\rm 1/2}$, peak velocity ($v$), and source angular dimension ($\theta$). 
The total column densities of all species have been evaluated assuming that the emission fills the telescope beam (i.e. no beam dilution has been applied), since we do not have any measurement of the emitting region in our sources. Therefore, the derived column densities are beam-averaged values. Observations towards low-mass protostars at high-angular resolution with the NOEMA interferometer indicate that the H$^{13}$CN and HC$^{15}$N(1--0) line emission is indeed very similar (Zapata et al.~\citeyear{zapata13}, Wampfler et al., personal comunication). Therefore, the assumed filling factor should not influence the $^{14}$N/$^{15}$N ratio.
We have fitted all the lines fixing also the {\it T}$_{\rm ex}$. We have used the excitation temperature derived from the K-ladder of the $J$=5-4 transition of the CH$_{3}$CN molecule, which is a good thermometer of massive dense cores (e.g. Araya et al.~\citeyear{araya05}, Purcell et al.~\citeyear{purcell06}). The detailed analysis of this species will be presented in a forthcoming paper (Mininni et al., in prep.).
The critical densities of the analysed transitions are relatively high ($\sim$3$\times$10$^{5}$ cm$^{-3}$), thus the observed transitions could be in non-LTE conditions. As shown by Daniel et al.~(\citeyear{daniel16}), if the excitation temperatures are lower than the kinetic temperatures, the excitation temperature of the less abundant  isotopologue (with $^{15}$N) could be different from that of the more abundant one and the isotopic ratio could also be affected. Therefore, we have checked with RADEX\footnote{ http://var.sron.nl/radex/radex.php} how T$_{ex}$ would change in non-LTE conditions and we have found that the T$_{ex}$ are lower comparing with those obtained with CH$_{3}$CN. However, the T$_{ex}$ of both isotopologues are similar (within $\sim$16\%), being that of the more abundant isotopologue always higher.
Therefore, the hypothesis that the rotational levels of the two molecules are populated with the same T$_{ex}$ is valid, and the column densities are computed with the same formula used in LTE approximation (Caselli \& Ceccarelli \citeyear{caselli12}). Furthermore, we have tested how the ratios can change if we use a non-LTE analysis (RADEX), and we have found that the ratios would be, on average, lower than a factor from 1.1 up to 1.6. However we stress that a precise non-LTE analysis can not be made because we don't know H$_{2}$ column densities and kinetic temperatures of all the sources.

After that, leaving free the other three parameters ({\it N}, {\it v} and $\Delta v_{\rm 1/2}$), we have used the AUTOFIT tool, which compares automatically the synthetic LTE spectra with the observed spectra, and provides the best non-linear least-squared fit using the Levenberg-Marquardt algorithm. Column densities, line widths, peak velocities and opacity ($\tau$) derived from the fit are given in Tables \ref{fit1} and \ref{fit2}. It can be seen that all the transitions are optically thin as expected. It should also be noted that the opacities are lower limits of the real opacity of the lines, since we assume that the emission of the lines comes from the entire beam. Despite this, we do not expect much higher opacities, and the transitions can be considered still optically thin. To test this, we have assumed a source size of 10'' and found that the $\tau$ values increase at most by a factor 10, so that the values would be $\sim0.01$--$0.1$. The assumption of optically thin transitions is hence still valid.
 
 In Table \ref{colden} column densities of the main isotopologues and of the corresponding $^{15}$N-bearing species are shown. 


\begin{figure*}
\centering
\includegraphics[width=40pc]{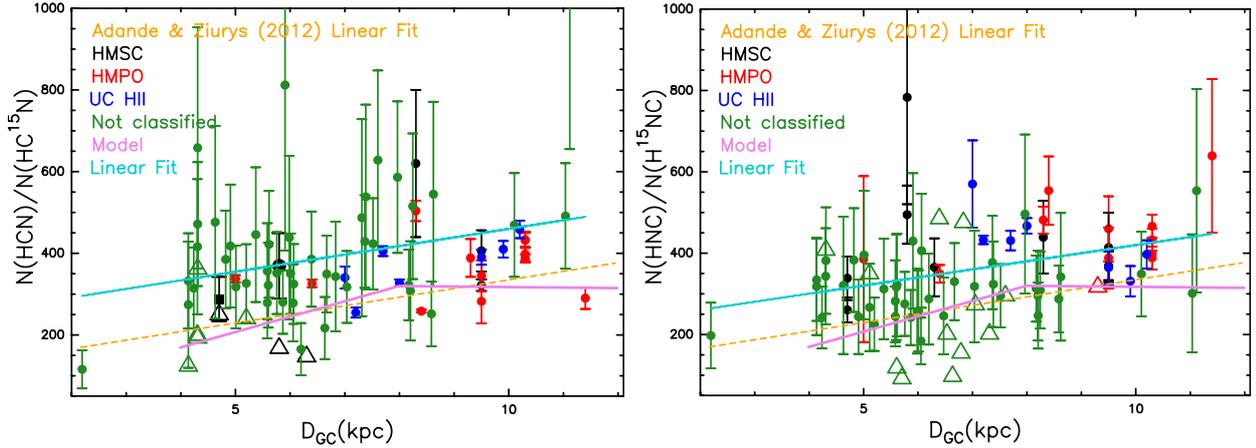}
\caption{$^{14}$N/$^{15}$N ratios for HCN (left panel) and HNC (right panel) as a function of Galactocentric distances of the sources listed in Table \ref{tab-coo}. Symbols are the same of Fig.~\ref{fig-colden} but the open triangles are lower limits for the corresponding $^{14}$N/$^{15}$N ratio. The cyan solid line is the linear regression fit computed for the two data sets and the pink solid line is the model of Romano et al.~(\citeyear{romano17}). The dashed orange line is the regression linear fit obtained by Adande \& Ziurys (\citeyear{adande12}). Note that we used the scale from 50 to 1000 of y-axes for visualization purposes, and that one source for HCN, with value 1305$\pm$650, falls outside.}
\centering
\label{fig-ratio}
\end{figure*}

\subsection{$^{14}$N/$^{15}$N ratios}
The reason why we have observed H$^{13}$CN and HN$^{13}$C is that HNC and HCN are usually optically thick in their lower energy rotational transitions (e.g. Padovani et al.~\citeyear{padovani11}). Hence, we decided to use the $^{13}$C-bearing species that have been found to be optically thin in high-mass star-forming cores with similar densities (Colzi et al.~\citeyear{colzi18}), and then, the $^{14}$N/$^{15}$N ratio was computed from the column density ratio of the two isotopologues and corrected by the $^{12}$C/$^{13}$C Galactic trend given in Milam et al.~(\citeyear{milam05}):
\begin{equation}
^{12}\textrm{C}/^{13}\textrm{C}=(6.01\pm1.19) \textrm{ kpc}^{-1}\times\textrm{D}_{\rm GC} +(12.28\pm9.33).
\label{milam}
\end{equation}
This trend is measured with observations of CN, CO and H$_{2}$CO in a sample of 18 molecular clouds that lie in the range of Galactocentric distances 0.09--16.41 kpc.
Note that $^{13}$C-fractionation may affect the abundances of $^{13}$C-bearing species. In fact nitriles and isonitriles are predicted to be significantly depleted in $^{13}$C (Roueff et al.~\citeyear{roueff15}). However, this depletion is at most of a factor 2 and is derived from a chemical model with a fixed kinetic temperature of 10 K, which probably is not the average kinetic temperature of our sources (see Table \ref{tab-coo}). The predictions of this model may therefore not be appropriate for our objects and observational tests to verify whether this theoretical effect is real have yet to be performed.

We have computed the $^{14}$N/$^{15}$N ratios for HNC and HCN, along with uncertainties derived propagating the error on total column densities (Table \ref{colden}). The uncertainties do not include the calibration errors, which cancels out in the ratio because the two lines were obtained in the same spectral setup. In Table \ref{colden} the $^{14}$N/$^{15}$N ratios are given.

As discussed in Sect.~\ref{observations}, we have implemented these observations with those made by Colzi et al.~(\citeyear{colzi18}). The observations of this sample were made with the same receiver and spectrometer of this work, giving us the possibility to combine the two samples with consistency. For further information about the observations and the results concerning the sample of 27 sources, see Colzi et al.~(\citeyear{colzi18}).
The isotopic ratios $^{14}$N/$^{15}$N, of the whole sample are in the range $\sim$115--1305 from HCN, and in the range $\sim$185--780 from HNC, as can be noted in Table \ref{colden}. All of the ratios found in this work are on average consistent with those computed by Colzi et al.~(\citeyear{colzi18}) (Fig.~\ref{fig-colden}). The most direct result of this work, based on an unprecedented statistics, is a firm confirmation of the finding suggested by Colzi et al.~(\citeyear{colzi18}): the lowest values found in pristine Solar System material (e.g. $\sim$150 in nitrogen hydrides of some comets, Manfroid et al.~\citeyear{manfroid09}) are not typical at the core scales.

In Fig.~\ref{fig-colden}, where the column densities of the two samples are shown, at the bottom right of each panel we have made a histogram, with bin width of 40, that shows the distribution on the ratios. We can note that for both molecules the distribution is asymmetric and centered in the bin 310$\leq^{14}$N/$^{15}$N$\leq$350, namely below the PSN value ($\sim$441) and just above the TA value ($\sim$272). 

\begin{figure*}
\centering
\includegraphics[width=43pc]{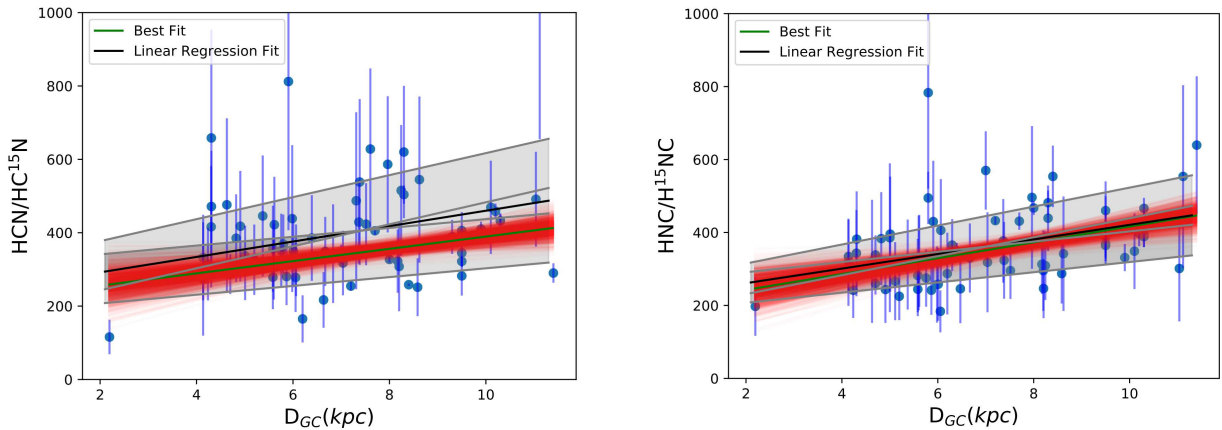}
\caption{$^{14}$N/$^{15}$N ratios for HCN (left panel) and HNC (right panel) as a function of Galactocentric distances of the sources. In all panels only the detection (not the upper limits), are represented with blue filled circles. The red solid lines are all the linear fit found with the Bayesian method, described in text, and the green solid line is the best fit performed with \emph{linmix}. For comparison we have added with black solid lines the linear regression fits (in grey are represented the errors on this fit), computed as explained in Sect.~\ref{res}. Note that we used the scale from 50 to 1000 of y-axes for visualization purposes, and that one source for HCN, with value 1305$\pm$650, falls outside.}
\centering
\label{fig-bayes}
\end{figure*}

\begin{figure*}
\centering
\includegraphics[width=40pc]{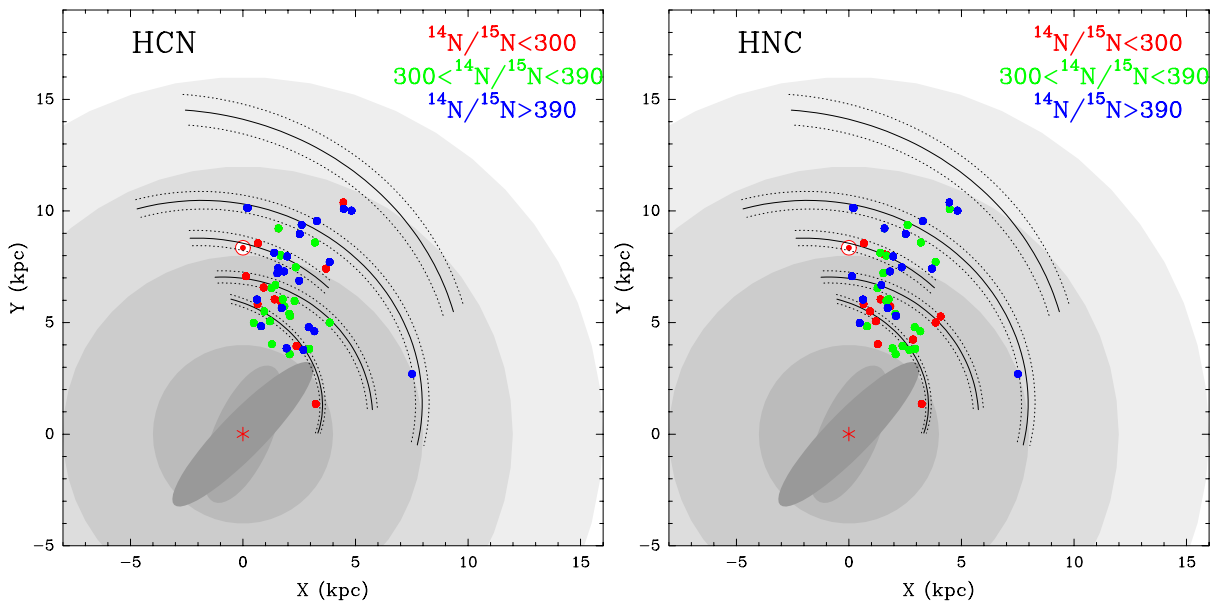}
\caption{Plan view of the Milky Way. The Galactic center (red asterisk) is at (0,0) and the Sun (red Sun symbol) is at (0,8.34). The background grey discs correspond to the Galactic bar region ($\sim$4 kpc), the solar circle ($\sim$8 kpc), co-rotation of spiral pattern ($\sim$12 kpc) and the edge of major star-formation regions ($\sim$16 kpc). The solid black lines are the center of spiral arms traced by masers, and the dotted lines the 1$\sigma$ widths. For more details see Reid et al.~(\citeyear{reid14}). The filled circles represent sources of our sample implemented with those of Colzi et al.~(\citeyear{colzi18}), and the three colors are the $^{14}$N/$^{15}$N ratios measured for HCN (left panel) and HNC (right panel): in red $^{14}$N/$^{15}$N$\le$300, in green 300$\le^{14}$N/$^{15}$N$\le$390 and in blue $^{14}$N/$^{15}$N$\ge$390.}
\centering
\label{fig-mappe}
\end{figure*}

\subsection{Galactocentric behavior: linear analysis}
The merged sample of 87 objects, given by the sources studied in this work and those analysed by Colzi et al.~(\citeyear{colzi18}), are located at different distances from the Galactic Centre ({\it D}$_{\rm GC}$). This gives us the opportunity to study the Galactocentric trend of the $^{14}$N/$^{15}$N ratio. 

Besides our study, one of the most recent works on the $^{14}$N/$^{15}$N Galactocentric gradient is that of Adande \& Ziurys (\citeyear{adande12}). These authors evaluated the ratio from rotational lines of CN, HCN and HNC towards a sample of mixed low- and high-mass star-forming regions observed with different telescopes. In total, their statistics is poorer, as the sample only contains 22 objects. They found an increasing trend of the $^{14}$N/$^{15}$N with {\it D}$_{\rm GC}$ and with an unweighted linear regression fit to the data set they obtained:
\begin{equation}
^{14}\textrm{N}/^{15}\textrm{N}=(21.1\pm5.2)\textrm{ kpc}^{-1}\times\textrm{D}_{\rm GC} +(123.8\pm37.1).
\end{equation}
Our total sample increases the statistics by a factor 4, and we can properly constrain for the first time the $^{14}$N/$^{15}$N Galactocentric trend with this robust statistics. We have chosen to separate the results of the two molecules because the distributions presented in Fig.~\ref{fig-colden} suggest a possible difference between the two isotopologues. We have first computed an unweighted linear regression fit to the data (excluding the lower limits), in order to compare it to the results of Adande \& Ziurys (\citeyear{adande12}), and found:
\begin{equation}
\textrm{HCN}/\textrm{HC}^{15}\textrm{N}=(21\pm9)\textrm{ kpc}^{-1}\times\textrm{D}_{\rm GC} +(250\pm67),
\end{equation}
\begin{equation}
\textrm{HNC}/\textrm{H}^{15}\textrm{NC}=(20\pm6)\textrm{ kpc}^{-1}\times\textrm{D}_{\rm GC} +(221\pm42).
\end{equation}
These trend are shown in Fig.~\ref{fig-ratio}.
Both slopes are consistent with that found by Adande \& Ziurys (\citeyear{adande12}), and are also consistent with each other. However, we found a general trend that shows an offset of about 100 toward larger values for the $^{14}$N/$^{15}$N, as can be noted also in both panels of Fig.~\ref{fig-ratio}. 
We have also tried to fit the data with a single power-law relation using the maximum-likelihood Bayesian tool \emph{linmix} by Kelly et al.~(\citeyear{kelly07}), which derives the linear dependence considering measurement uncertainties (Fig.~\ref{fig-bayes}) and excluding the lower limits. The best fits we found are:
\begin{equation}
\label{bay1}
\textrm{HCN}/\textrm{HC}^{15}\textrm{N}=(17\pm6) \textrm{ kpc}^{-1}\times\textrm{D}_{\rm GC} +(223\pm12),
\end{equation}
\begin{equation}
\label{bay2}
\textrm{HNC}/\textrm{H}^{15}\textrm{NC}=(22\pm6) \textrm{ kpc}^{-1}\times\textrm{D}_{\rm GC} +(198\pm12).
\end{equation}
The linear regression fit is consistent with the best Bayesian fit within the uncertainties, as can be seen in Fig.~\ref{fig-bayes}, mostly for HNC. Note that in the case of the Bayesian analysis the errors are lower than in the linear regression fit because in the standard analysis the errors on the parameters are computed separately, as they are considered independent. On the contrary in the Bayesian analysis an optimised procedure, that takes into account the dependence between the two parameters, is performed and converges to the best parameters with lower errors than in the other analysis.

\subsubsection{Dependence with spatial distribution in the Galaxy}
In order to investigate a dependence with the Galactic longitude ($l$, listed in Table \ref{tab-coo}), we have plotted our results in the Galactic plan view of the Milky Way made by Reid et al.~(\citeyear{reid14}). The solid curved lines trace the center of spiral arms as measured with masers associated with young, high-mass stars. We have plotted in the Galactic plane the position of our sources and we have divided the ratio in three intervals: $^{14}$N/$^{15}$N$\le$300, 300$\le$ $^{14}$N/$^{15}$N$\le$390 and $^{14}$N/$^{15}$N$\ge$390 (Fig.~\ref{fig-mappe}). We have decided to use these ranges because each of them contains a similar number of objects. For both molecules no clear trend along spiral arms was found. We have also checked if there was any residual trend after substracting to all the $^{14}$N/$^{15}$N ratios the values obtained at the correponding {\it D}$_{\rm GC}$ of the source with the Bayesian fit, but we have not found trends either. Thus, we conclude that the Galactocentric gradient, probably, is related only to the source distances and not to other local processes.

\begin{figure*}
\centering
\includegraphics[width=40pc]{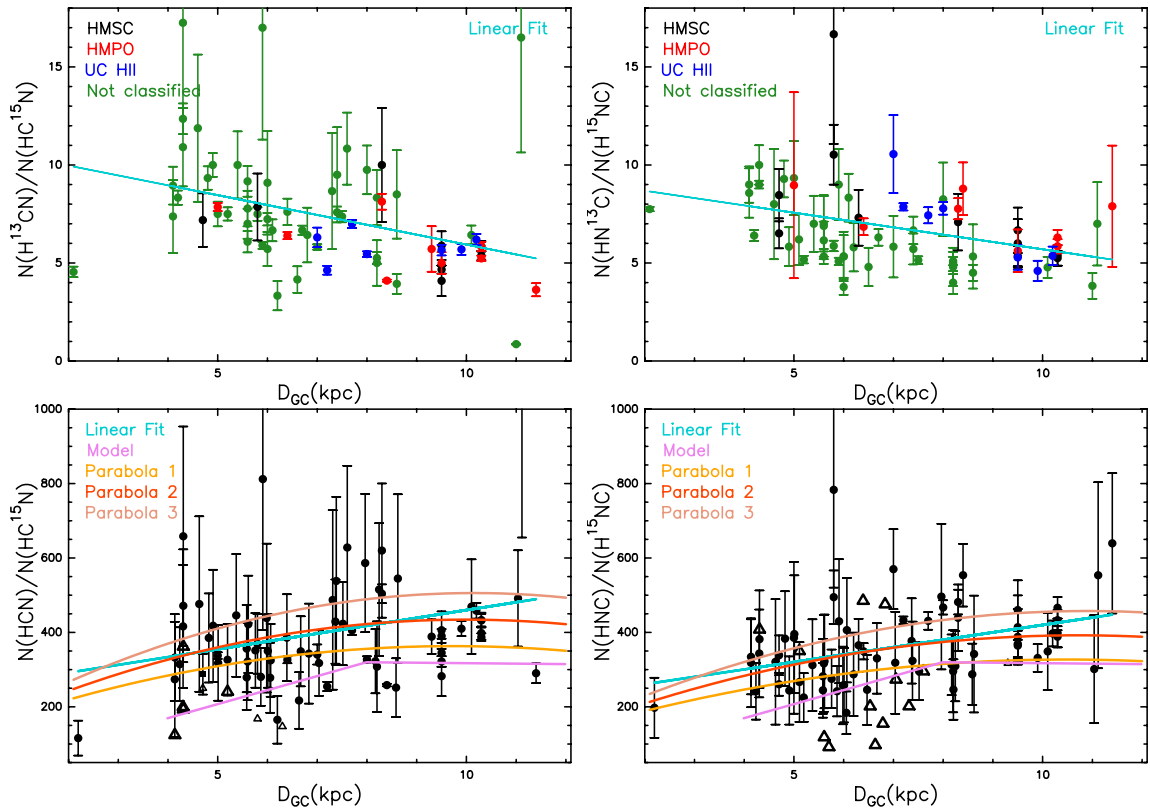}
\caption{Upper panels: $^{13}$C$^{14}$N/$^{12}$C$^{15}$N ratios for HCN (left panel) and HNC (right panel) as a function of Galactocentric distances of the sources. Symbols are the same of Fig.~\ref{fig-ratio} and the cyan solid line is the linear regression fit computed from the two data sets. Note that the negative slope is the consequence of the two combined ratios, $^{13}$C/$^{12}$C and $^{14}$N/$^{15}$N. Bottom panels: same of Fig.~\ref{fig-ratio} with all the sources in black, while the three parabolas are described in Sect.~\ref{res-par}. The yellow one is found using the minimum value in slope of Eq.~\eqref{milam} (Parabola 1), the red one is found using the central value of the slope (Parabola 2) and the pink one is found using the maximum value of the slope (Parabola 3). Note that we used the scale from 50 to 1000 of y-axes for visualization purposes, and that one source for HCN, with value 1305$\pm$650, falls outside.}
\centering
\label{fig-par}
\end{figure*}

\subsection{Galactocentric behavior: parabolic analysis}
\label{res-par}
The linear analysis relies on the $^{12}$C/$^{13}$C relation by Milam et al.~(\citeyear{milam05}) (Eq.~\eqref{milam}),
that we have assumed to compute $^{14}$N/$^{15}$N ratios.
Note that what can be measured directly from our observations are the H$^{13}$CN/HC$^{15}$N and HN$^{13}$C/H$^{15}$NC ratios. In the upper panels of Fig.~\ref{fig-par} these ratios have been plotted as a function of the Galactocentric distances, and, since the points suggest a negative linear trend, we have computed unweighted linear regression fits to these data, obtaining:
\begin{equation}
\textrm{H}^{13}\textrm{CN}/\textrm{HC}^{15}\textrm{N}=(-0.5\pm0.2) \textrm{ kpc}^{-1}\times\textrm{D}_{\rm GC} +(11.0\pm1.2),
\end{equation}
\begin{equation}
\textrm{HN}^{13}\textrm{C}/\textrm{H}^{15}\textrm{NC}=(-0.4\pm0.1) \textrm{ kpc}^{-1}\times\textrm{D}_{\rm GC} +(9.4\pm0.8).
\end{equation}
We have tested the linearity of this relation also using a non-parametric statistics that allows to explore the data sets without assuming an underlying model that describes the data, finding a linear trend of the ratios with the Galactocentric distances.
The $^{14}$N/$^{15}$N ratios, that have been derived in this work, are the product between $^{14}$N$^{13}$C/$^{15}$N$^{12}$C and $^{12}$C/$^{13}$C. Considering that $^{14}$N$^{13}$C/$^{15}$N$^{12}$C ratios are well fitted by a linear trend, and the assumption that $^{12}$C/$^{13}$C is also linear with {\it D}$_{\rm GC}$ (Milam et al.~\citeyear{milam05}), then the derived $^{14}$N/$^{15}$N must have a parabolic one. 
The parabolas have equations, for HNC and HCN, respectively:
\begin{equation}
\begin{split}
&\textrm{HCN}/\textrm{HC}^{15}\textrm{N}=[(-0.5\pm0.2) \textrm{ kpc}^{-1}\times\textrm{D}_{\rm GC} +(11.0\pm1.2)]\times\\
&\times[^{12}\textrm{C}/^{13}\textrm{C}],
\end{split}
\end{equation}
\begin{equation}
\begin{split}
&\textrm{HNC}/\textrm{H}^{15}\textrm{NC}=[(-0.4\pm0.1) \textrm{ kpc}^{-1}\times\textrm{D}_{\rm GC} +(9.4\pm0.8)]\times\\
&\times[^{12}\textrm{C}/^{13}\textrm{C}].
\end{split}
\end{equation}
In the bottom panels of Fig.~\ref{fig-par} the three parabolas obtained choosing the slope of $^{12}$C/$^{13}$C as the minimum, the centered and the maximum in the error, are plotted.
Interestingly, with this analysis we are able to reproduce the flattening trend above $\sim$8 kpc predicted by the GCE model of Romano et al.~(\citeyear{romano17}). This is the distance where the parabolic trend start to disagree with the linear trend of the linear analysis. 
This result remains the same whatever the assumption of the $^{12}$C/$^{13}$C.

\section{Discussion and Conclusions} 
\label{disc}
We have observed the rotational transitions \emph{J}=1-0 of HN$^{13}$C, H$^{15}$NC, H$^{13}$CN and HC$^{15}$N towards 66 massive star-forming cores to derive the $^{14}$N/$^{15}$N ratios. We have complemented this sample with that observed by Colzi et al.~(\citeyear{colzi18}), obtaining a total sample of 87 sources. The isotopic ratios measured range from 115 up to 1305 for HCN and from 185 up to 780 for HNC, which are higher than the low values in the pristine Solar System material. 

With the strong statistics of 87 sources a new Galactocentric trend has been derived. From the linear fits given in Eq.~\eqref{bay1} and \eqref{bay2}, we derive that the $^{14}$N/$^{15}$N ratios in the local interstellar medium (ISM), i.e. the values calculated at the Galactocentric distance of the Sun (8.4 kpc, Reid et al.~\citeyear{reid09}), are 383$\pm$51 for HCN and 366$\pm$51 for HNC. These values are both higher than 301$\pm$44 which was obtained from the gradient of Adande \& Ziurys (\citeyear{adande12}). Despite the large uncertainties, our new local interstellar values are more similar to the PSN value of $\sim$441, measured from the Solar wind adopting a totally different approach (Marty et al.~\citeyear{marty10}).
  
Let us now discuss briefly our findings in the general context of the origin of $^{14}$N and $^{15}$N in the Galaxy. As discussed in Sect.\ref{intro}, primary contributors to $^{14}$N are massive stars in He-shell burning, while the main contributor of $^{15}$N, which is a secondary element, is the hot CNO cycle in novae outbursts. In our Galaxy the abundance of heavy elements is found to decrease with the distance, as observed by Maciel \& Costa (\citeyear{maciel09}). Therefore the $^{14}$N/$^{15}$N ratio should increase with Galactocentric distance as the main contributors of $^{15}$N are novae and some supernovae events (for example, Romano et al.~\citeyear{romano17}). The positive trend found with the Galactocentric distance suggests that there is not a primary component production of $^{15}$N, that is instead important for $^{14}$N in the environment with an high rate of star formation, like at {\it D}$_{\rm GC}$<8 kpc. 
The slope of the GCE model of Romano et al.~(\citeyear{romano17}), which takes novae as the main contributors of $^{15}$N into account, is in agreement with our observational results (Fig.~\ref{fig-ratio}), up to 8 kpc. Then, up to 12 kpc the curve in the model flattens. With our linear fit we can not say what is happening after 8 kpc as the linear trend can be biased by points at smaller distances. 

A second analysis points out a parabolic trend of $^{14}$N/$^{15}$N ratios with the Galactocentric distance. This is based taking into account the linear assumption made for the $^{12}$C/$^{13}$C ratios (Milam et al.~\citeyear{milam05}). With this analysis we were able to reproduce the flattening trend above 8 kpc, as predicted by the GCE model of Romano et al.~(\citeyear{romano17}).
The flattening of the $^{14}$N/$^{15}$N gradient for {\it D}$_{\rm GC}>$ 8 kpc in the model of Romano et al.~(\citeyear{romano17}) is due to the absence of substantial $^{14}$N-enrichment from low-metallicity massive stars (Nomoto et al.~\citeyear{nomoto13}).

Note that the model is a lower limit of the $^{14}$N/$^{15}$N ratios found (Fig.~\ref{fig-ratio} and bottom panels of Fig.~\ref{fig-par}). An interpretation of this result could be that the first N-fractionation is regulated by the nucleosynthesis processes, and afterwards there could be local (at core level) enrichments of $^{14}$N/$^{15}$N.

Of course, to confirm our results, more observations of sources in the outer Galaxy are needed. New observations could place important constraints on the amount of $^{14}$N produced by massive stars in low-metallicity environments. 
The trend we have found in our Milky Way can be relevant
also as "template" for extra-galactic sources, in which the $^{14}$N/$^{15}$N
has been measured in a few objects. For example, toward the massive star forming region N113 in the Large Magellanic Cloud a $^{14}$N/$^{15}$N ratio of 111$\pm$17 was found (Chin et al.~\citeyear{chin99}). Toward the nuclear region of nearby Seyfert 2 galaxy NGC 1068 a $^{14}$N/$^{15}$N ratio greater than 419 was found (Wang et al.~\citeyear{wang14}). Toward the z=0.89 spiral galaxy, located on the line of sight to the quasar PKS1830-211, a ratio of 152$\pm$27 was measured (Muller et al.~\citeyear{muller11}). Moreover, through millimeter molecular absorption lines arising in the z=0.68 absorber toward B0218+357, a ratio of $\sim$120 was measured (Wallström et al.~\citeyear{wallstrom16}). These measurements indicate that even in external galaxies the $^{14}$N/$^{15}$N is far from being homogeneous. We stress, however, that the comparison between local clouds in the Milky Way and external galaxies should be taken with caution. In this kind of sources the relative contribution of the processes responsible for changes in the $^{14}$N/$^{15}$N (nucleosynthesis and/or chemical fractionation), has not been explored. Moreover,
it is difficult to determine local trends because the single clouds cannot be resolved easily.

\section*{Acknowledgements}
We are grateful to the IRAM-30m telescope staff for their help during the observations. 
Many thanks to the anonymous referee for the careful reading of the paper and the comments that improved the work.
We would like to thank Mark Reid and the BeSSel Project for the plot of the structure of our Galaxy.
Paola Caselli acknowledges support from the European Research Council (project PALs 320620).
V. M. Rivilla has received funding from the European Union's Horizon 2020 research and innovation programme under the Marie Sk\l{}odowska-Curie grant agreement No 664931. \'A.\, S.-M. acknowledges support by the Deutsche Forschungsgemeinschaft (DFG) in the framework of the collaborative research grant SFB\,956. project A6.





{}
\onecolumn
\begin{longtable}{lccccccc}
\caption{\label{tab-coo} List of the observed sources. Cols. 2, 3, 4 and 5 give the coordinates (equatorial and Galactic) of the sources and Col. 6 give the source distance to the Sun of the associated star-forming region. In Col.~7 the Galactocentric distances of the sources obtained from the distance to the Sun and the corresponding Galactic longitude (Col.~4) are listed. In Col.~8 we give the {\it T}$_{\rm ex}$ used to fit the lines for all the sources, computed as explained in Sect.~\ref{res}. The {\it T}$_{\rm ex}$ have been derived fitting the K-ladder of CH$_{3}$CN(5--4) (Mininni et al., in prep.).}
  \tabularnewline \hline \hline
Source & $\alpha({\rm J2000})$ &$\delta({\rm J2000})$ & $l$ & $b$ & $d$ & {\it D}$_{GC}$ & {\it T}$_{\rm ex}$  \\
& (h m s)& ($\degr$ $\arcmin$ $\arcsec$) & ($\degr$)  &($\degr$) & (kpc) &(kpc) & (K)\\
\hline
\endfirsthead
\caption{Continued.}\\ 
\hline \hline
Source & $\alpha({\rm J2000})$ &$\delta({\rm J2000})$ & $l$ & $b$ & $d$ & {\it D}$_{GC}$ & {\it T}$_{\rm ex}$  \\
& (h m s)& ($\degr$ $\arcmin$ $\arcsec$) & ($\degr$)  &($\degr$) & (kpc) &(kpc) & (K)\\
\hline
 \endhead
I00117-MM1$^{**}$     & 00:14:26.1 & +64:28:44.0    &  118.96 & +1.89          &   1.8 & 9.4& 35$^{***}$ \\
I04579-VLA1$^{**}$   & 05:01:39.9 & +47:07:21.0   &    160.14 & +3.16            &   2.5  &10.7   & 35$^{***}$ \\
18089-1732M1   & 18:11:51.5 & -17:31:29.0   &   12.89 & +0.49           &    3.6* &4.9   & 87 \\
18089-1732M4   & 18:11:54.0 & -17:29:59.0   &   12.92 & +0.49           &    3.6* &4.9  & 36\\
18151-1208M1   & 18:17:58.0 & -12:07:27.0   &   18.34 & +1.77           &    3.0  &5.6  & 37\\
18151-1208M2   & 18:17:50.4 & -12:07:55.0   &   18.32 & +1.79           &    3.0  & 5.6 & 42\\
18182-1433M1   & 18:21:09.2 & -14:31:49.0   &   16.59 & -0.05           &    4.5* &4.2  & 52 \\
18264-1152M1   & 18:29:14.6 & -11:50:22.0   &   19.88 & -0.53           &    3.5* & 5.2 & 50 \\
18272-1217M1   & 18:30:02.9 & -12:15:17.0   &   19.61 & -0.90                &     2.9    & 5.7 & 52  \\
18290-0924M2   & 18:31:46.3 & -09:22:23.0   &   22.36 & +0.06           &    5.3* & 4.0& -- \\
18306-0835M1   & 18:33:24.0 & -08:33:31.0   &   23.27 & +0.08           &    4.9* & 4.3& 39 \\
18306-0835M2   & 18:33:17.2 & -08:33:26.0   &   23.25 & +0.10           &    4.9* & 4.3& 36 \\
18308-0841M1   & 18:33:33.1 & -08:39:10.0   &   23.20 & -0.00           &    4.9* & 4.3 & 36\\
18310-0825M2   & 18:33:44.0 & -08:21:20.0   &   23.48 & +0.10           &    5.2* & 4.1& 38\\
18310-0825M3   & 18:33:42.2 & -08:21:36.0   &   23.48 & +0.10           &    5.2* & 4.1& 34\\
18372-0541M1   & 18:39:56.0 & -05:38:47.0   &   26.60 & -0.02           &    1.8* &  6.8& 71\\
18385-0512M1   & 18:41:13.3 & -05:09:01.0   &   27.19 & -0.08           &    2.0* &  6.6& 49\\
18445-0222M3   & 18:47:07.7 & -02:21:26.0   &   30.34 & -0.12           &    5.3* &  4.6& 15\\
18454-0136M1   & 18:48:02.5 & -01:33:26.0   &   31.16 & +0.05           &    2.7* & 6.2& 33\\
18472-0022M1   & 18:49:52.4 & -00:18:56.0   &   32.47 & +0.20           &    3.2* &  5.9& 32\\
18488+0000M1   & 18:51:25.6 & +00:04:06.0   &   32.99 & +0.03           &    5.4* & 4.8& 36\\
18517+0437M1   & 18:54:14.3 & +04:41:40.0   &   37.43 & +1.5           &     2.9& 6.3& 41\\
18521+0134M1   & 18:54:40.8 & +01:38:05.0   &   34.76 & +0.02           &    5.0* & 5.1& 37\\
19012+0536M1   & 19:03:45.4 & +05:40:43.0   &   39.39 & -0.14           &    4.6* & 5.6& 37\\
19035+0641M1   & 19:06:01.6 & +06:46:43.0   &   40.62 & -0.14           &    2.2 & 6.8 & 49\\
19095+0930    & 19:11:54.0 & +09:35:52.0   &     43.80 & -0.13            &    3.3 &6.4 & 57\\
19282+1814M1   & 19:30:23.1 & +18:20:25.0   &     53.62 & +0.03           &    1.9* & 7.4& 44\\
19410+2336M1   & 19:43:10.9 & +23:44:10.0   &     59.78 & +0.07          &    2.1* &7.5 & 32\\
19411+2306M1   & 19:43:18.0 & +23:13:59.0   &     59.36 & -0.21           &   2.9* & 7.3  & 27 \\
19413+2332M1   & 19:43:29.3 & +23:40:18.0   &     59.76 & -0.03           &   1.8* &  7.6 & 36 \\
ON1$^{**}$            & 20:10:09.1 & +31:31:36.0   &     69.54 & -0.98           &   2.5 & 7.8  & 85 \\
20126+4104M1   & 20:14:25.9 & +41:13:34.0   &     78.12 & +3.63           &   1.7 &  8.2 & 71 \\
20216+4107M1   & 20:23:23.5 & +41:17:38.0   &     79.13 & +2.28           &   1.7 & 8.2& 54 \\
20319+3958M1   & 20:33:49.1 & +40:08:35.0   &     79.35 & +0.00           &   1.6 & 8.2 & 58 \\
20332+4124M1   & 20:34:59.7 & +41:34:49.0   &     80.63 & +0.68           &   3.9 & 8.6& 35$^{***}$ \\
20343+4129M1   & 20:36:07.9 & +41:39:57.0   &     80.83 & +0.57           &   1.4 &  8.2& 52 \\
22187+5559V3$^{**}$   & 22:20:33.7 & +56:14:29.1   &    103.06 & -0.69           &   2.9 & 9.4 & --\\
22187+5559V5$^{**}$   & 22:20:35.6 & +56:14:46.4   &    103.06 & -0.69           &   2.9 & 9.4 & --\\
22198+6336$^{**}$    & 22:21:26.8 & +63:51:37.2   &    107.30 & +5.64           &     0.7& 8.6& 64 \\
22570+5912M2   & 22:58:59.2 & +59:27:36.0   &   109.08 & -0.35            &   5.1 &  11.1& 35$^{***}$  \\
23033+5951M1   & 23:05:25.3 & +60:08:06.0   &   110.09 & -0.07            &   3.5 & 10.1 & 26  \\
23139+5939M1   & 23:16:10.5 & +59:55:28.0   &   111.26 & -0.77            &   4.8 & 11.0 & 39  \\
G75-core$^{**}$     & 20:21:44.0 & +37:26:38.0   &    75.78 & +0.34            &   3.8 &8.3 & 56 \\
G08.14+0.22$^{**}$  & 18:03:01.3 & -21:48:05.0   &      8.14 & +0.22            &   3.4  & 5.0& 52 \\
G14.33-0.65$^{**}$  & 18:18:54.8 & -16:47:53.0   &     14.33 & -0.65            &    2.6 & 5.9 & 49 \\
G14.99-0.67$^{**}$  & 18:20:17.6 & -16:13:55.0   &     14.99 & -0.67  &    2.4  & 6.0 & 22\\
G14.99-2$^{**}$      & 18:20:17.6 & -16:13:55.0  &     14.99 & -0.67            &    2.4 &6.0  & 35$^{***}$\\
G15.02-0.62$^{**}$  & 18:20:10.3 & -16:10:35.0  &     15.02 & -0.62           &   2.4 & 6.1 & 14\\
G24.78+0.08$^{**}$  & 18:36:12.6 & -07:12:10.9  &     24.79 & +0.08           &   7.7 & 2.1 & 93\\
G31.41+0.31$^{**}$  & 18:47:34.2 & -01:12:45.0  &     31.41 & +0.31           &   7.9 &3.5 & --\\
G34.82+0.35$^{**}$  & 18:53:37.4 & +01:50:32.0  &     34.82 & +0.35           &   3.6 & 5.6 & 35\\
G35.03+0.35$^{**}$  & 18:54:00.6 & +02:01:19.3  &     35.02 & +0.35           &   3.2 & 6.0 & 40\\
G35.20-0.74$^{**}$  & 18:58:13.0 & +01:40:36.0  &     35.20 & -0.74           &   2.2 & 6.7  & 50\\
G36.70+0.09$^{**}$  & 18:57:59.3 & +03:24:05.0  &     36.71 & +0.10           &   9.7 & 5.6 & --\\
G37.55+0.19$^{**}$  & 18:59:11.4 & +04:12:14.0  &     37.56 & +0.20           &   5.6 &5.1  & 44\\
G40.28-0.22$^{**}$  & 19:05:42.1 & +06:26:08.0  &     40.28 & -0.22           &   4.9 & 5.4 & 43\\
G42.03+0.19$^{**}$  & 19:07:29.0 & +08:10:39.0  &     42.03 & +0.19           &  11.1 & 7.2 & --\\
G42.30-0.30$^{**}$  & 19:09:44.2 & +08:11:33.0  &     42.30 & -0.30           &  10.5 & 7.0 & 35$^{***}$\\
G42.70-0.15$^{**}$  & 19:09:55.8 & +08:36:56.0  &     42.70 & -0.15           &  15.9 &11.2 & --\\
G48.99-0.30$^{**}$  & 19:22:26.3 & +14:06:37.0  &     48.99 & -0.30           &   5.1 & 5.8 & 37\\
G49.41+0.33$^{**}$  & 19:20:58.9 & +14:46:46.0  &     49.41 & +0.33           &  12.2 &9.2  & --\\
G50.78+0.15$^{**}$  & 19:24:17.2 & +15:53:54.0  &     50.78 & +0.15           &   7.0 & 6.5 & --\\
G52.92+0.41$^{**}$  & 19:27:35.2 & +17:54:26.0  &     52.92 & +0.41           &   5.1 & 6.5 & --\\
G53.04+0.11$^{**}$  & 19:28:55.7 & +17:52:01.0  &     53.04 & +0.11           &   9.4 &8.0  & 32\\
G53.14+0.07$^{**}$  & 19:29:17.5 & +17:56:24.0  &     53.14 & +0.07           &    1.9  & 7.4 & 42\\
NGC7538-IRS1$^{**}$  & 23:13:43.3&  +61:28:10.6  &   111.54 & +0.78            &    2.8 &7.0  & 43 \\
NGC7538-IRS9$^{**}$  & 23:14:01.8 & +61:27:20.0   &    111.57 & +0.75            &   2.8&9.7 & 39 \\
 \bottomrule
 \end{longtable}
* sources with distance ambiguity, due to their posiition within the Solar circle, for which we have chosen the near distance. For details see Sridharan et al.~(\citeyear{sridharan02}).\\ 
$^{**}$ sources selected from the literature (Fontani et al.~\citeyear{fontani11}, Wouterloot et al.~\citeyear{wouterloot93}, Beltrán et al.~\citeyear{beltran16}, Cesaroni et al.~\citeyear{cesaroni17}, Tan et al.~\citeyear{tan14}).\\
$^{***}$ sources for which it was not possible to obtain {\it T}$_{\rm ex}$ from CH$_{3}$CN and then we have used a {\it T}$_{\rm ex}$ of 35 K, which is the mean value of the excitation temperatures of the sources.

\newpage
\begin{longtable}{lllllllc}
\caption{\label{colden} Total column densities of observed transitions, and $^{14}$N/$^{15}$N ratios. In Cols.~2, 3, 4 and 5 are listed the total column densities (beam-averaged), computed as explained in Sect.~\ref{res}, of HN$^{13}$C, H$^{15}$NC, H$^{13}$CN and HC$^{15}$N. In Col.~6 and 7 the $^{14}$N/$^{15}$N ratios for HNC and HCN are listed.  In Col.~8 the$^{12}$C/$^{13}$C ratio, used to compute the ratio, are listed.}
\tabularnewline \hline \hline
Source & N(HN$^{13}$C) & N(H$^{15}$NC) &N(H$^{13}$CN) & N(HC$^{15}$N)& $\frac{HNC}{H^{15}NC}$ & $\frac{HCN}{HC^{15}N}$ & $^{12}$C/$^{13}$C \\
& ($\times$10$^{12}$cm$^{-2}$) & ($\times$10$^{11}$cm$^{-2}$) &($\times$10$^{12}$cm$^{-2}$) & ($\times$10$^{11}$cm$^{-2}$)& & & \\
\hline
\endfirsthead
\caption{Continued.} \\
\hline \hline
Source & N(HN$^{13}$C) & N(H$^{15}$NC) &N(H$^{13}$CN) & N(HC$^{15}$N)& $\frac{HNC}{H^{15}NC}$ & $\frac{HCN}{HC^{15}N}$ & $^{12}$C/$^{13}$C  \\
& ($\times$10$^{12}$cm$^{-2}$) & ($\times$10$^{11}$cm$^{-2}$) &($\times$10$^{12}$cm$^{-2}$) & ($\times$10$^{11}$cm$^{-2}$)& & \\
\hline
\endhead
I00117-MM1   & 1.5$\pm$ 0.3       &       $\le$5.9&     2.7    $\pm$   0.4          &      5      $\pm$  2 $^{t}$                   &  $\ge$174 &  370     $\pm$   139         & 68   \\      
I04579-VLA1  & $\le$1.1     &       $\le$ 10.1&    4.2$\pm$0.8  &    $\le$11 &  --&  $\ge$293      &  77 \\
18089-1732M1 &  35    $\pm$  4          &         47    $\pm$   8     &  78   $\pm$    9        &      140   $\pm$   20      &311     $\pm$  97   &  233     $\pm$   68  &   42    \\ 
18089-1732M4 &    3.5  $\pm$    0.5          &     6    $\pm$     2       &    7.0   $\pm$    1.0       &    7     $\pm$   2 & 244     $\pm$   92   &  418   $\pm$   151      &   42    \\       
18151-1208M1 &    8    $\pm$    1         &        15   $\pm$    3        &   14   $\pm$    2         &      23   $\pm$   4  & 244   $\pm$   74  &  279    $\pm$   88    &   46    \\      
18151-1208M2 &    8    $\pm$    1          &       13   $\pm$    3        &  14   $\pm$    2         &       18   $\pm$   4  & 282     $\pm$   101  &  356     $\pm$   118    &  46   \\      
18182-1433M1 &    16   $\pm$   2         &         25   $\pm$    4       &  40   $\pm$    4         &       48   $\pm$   6   & 241    $\pm$   76  &  314  $\pm$   99   &   38  \\      
18264-1152M1 &    15   $\pm$   2          &        29   $\pm$    4        &  36   $\pm$    4         &      48   $\pm$   7   & 225  $\pm$   66   &  326   $\pm$   96   &  43   \\       
18272-1217M1 &1.4$\pm$ 0.3      &       $\le$7.1&   4.5$\pm$0.7  &     $\le$9.3 & $\ge$92 &  $\ge$225    &  46  \\
18290-0924M2 &$\le$0.6       &     $\le$6.2 &      $\le$1  &  $\le$7.1   & --&  --       &  36  \\
18306-0835M1 &   10    $\pm$  1            &  10    $\pm$   2   &  21   $\pm$    3         &     17   $\pm$   3      & 382    $\pm$   131  &  471     $\pm$   152      &  38  \\     
18306-0835M2 &4.7$\pm$ 0.6       &      $\le$4.4            &  6.9   $\pm$    0.9         &     4     $\pm$   2     & $\ge$408 &  658    $\pm$   296           &  38   \\  
18308-0841M1 &   9     $\pm$   1            &      10    $\pm$   2         &     12   $\pm$    2        &      11   $\pm$   3  & 343     $\pm$   118  &  416    $\pm$   166      &   38 \\    
18310-0825M2 &   9     $\pm$   1         &         10    $\pm$   1        &  17   $\pm$    2         &      19   $\pm$   4     & 334     $\pm$   101   &  332     $\pm$   117      &  37   \\    
18310-0825M3 &   6.0   $\pm$   0.8          &      7     $\pm$    2        &       5.9   $\pm$    0.8     &    8     $\pm$   3$^{t}$ & 318     $\pm$   120   &  274     $\pm$   155       &   37  \\           
18372-0541M1 & 3.5$\pm$ 0.7      &      $\le$1.2&   12$\pm$1  &   $\le$13 & $\ge$155 &  $\ge$490        &   53   \\ 
18385-0512M1 & 1.7$\pm$ 0.4      &       $\le$9.1&  8   $\pm$    1         &     19   $\pm$   5     & $\ge$97&  217     $\pm$   76     &  52   \\ 
18445-0222M3 &   1.6   $\pm$  0.2           &      2.0   $\pm$   0.9       &    1.9   $\pm$    0.3        &     1.6   $\pm$   0.6  & 321     $\pm$   169   &  476     $\pm$   236    &   40   \\      
18454-0136M1 &   2.9   $\pm$  0.5         &        5     $\pm$   1       &  2.0   $\pm$    0.5         &      6     $\pm$   2      & 287     $\pm$   112  &  165    $\pm$   65      &  49  \\    
18472-0022M1 &   4.5   $\pm$  0.6          &       5     $\pm$   2        &  5.1   $\pm$    0.7         &     3     $\pm$   1      & 430     $\pm$   167   &  812     $\pm$   406    &   48   \\     
18488+0000M1 &   6.5   $\pm$  0.8          &       7     $\pm$   1        &  14   $\pm$    2         &       15   $\pm$   2    & 383   $\pm$   128   &  385     $\pm$   120     &  41  \\     
18517+0437M1 &   7.8   $\pm$  0.9            &     13    $\pm$  2          &   23   $\pm$    2         &      35   $\pm$   4   & 301     $\pm$   85   &  329     $\pm$   89     &  50  \\       
18521+0134M1 &   3.1   $\pm$  0.5            &     5     $\pm$   2         &   4.8$\pm$0.7  &    $\le$6.3& 267     $\pm$   107   & $\ge$328       &   43    \\
19012+0536M1 &   3.5   $\pm$  0.5            &     5     $\pm$   1$^{t}$      &    11   $\pm$    1         &      12   $\pm$   2 & 322     $\pm$   126  &  422     $\pm$   131     &  46    \\         
19035+0641M1 & 6.6$\pm$ 0.9      &       $\le$7.4            &      9     $\pm$    1                &       14   $\pm$   4 $^{t}$   & $\ge$476 &  343     $\pm$   134         &   53  \\ 
19095+0930   & 9$\pm$ 1      &       $\le$9.4       &  35   $\pm$    4         &     46   $\pm$   8     & $\ge$485  &  385     $\pm$   117     &       51    \\  
19282+1814M1 &   4.0   $\pm$   0.6          &     7     $\pm$    2          &      3.8   $\pm$    0.6      &     4     $\pm$   2  & 324     $\pm$   106   &  538    $\pm$   226    &   57     \\              
19410+2336M1 &   7.2   $\pm$   0.9          &     14    $\pm$   2           &      14   $\pm$    2         &     19   $\pm$   3   & 295     $\pm$   78   &  423     $\pm$   112      &   57   \\           
19411+2306M1 & 1.9$\pm$ 0.3      &     $\le$5.3         &  2.6   $\pm$    0.5         &      3     $\pm$   1    & $\ge$202&  487     $\pm$   241        &  56 \\  
19413+2332M1 & 2.6$\pm$ 0.4      &     $\le$5.1        &  6.5   $\pm$    0.8         &       6     $\pm$   2    & $\ge$296 &  628     $\pm$   220       &  58  \\ 
ON1          &    24   $\pm$   3         &        40    $\pm$   9            &      35   $\pm$    4         &     55   $\pm$   9  & 356    $\pm$   113  &  378    $\pm$   103   &   59    \\       
20126+4104M1 &    26   $\pm$   3        &         51    $\pm$   7            &    31   $\pm$    4         &       59   $\pm$   9  & 313     $\pm$   82   &  323     $\pm$   87    &  61   \\       
20216+4107M1 &    4.8  $\pm$    0.7        &      10    $\pm$     3          &       4.5   $\pm$    0.7     &     9     $\pm$   3 & 296     $\pm$   111   &  306     $\pm$   122     &  62   \\              
20319+3958M1 &    2.8  $\pm$    0.4        &      7     $\pm$    2           &  3.2$\pm$0.5        &  $\le$8.2 & 246     $\pm$   81   & $\ge$240   &   62    \\ 
20332+4124M1 &    1.6  $\pm$    0.3        &      3     $\pm$    1$^{t}$         &      3.4   $\pm$    0.5      &     4     $\pm$   1  & 342     $\pm$   157   &  545     $\pm$   227     &  64  \\            
20343+4129M1 &    9    $\pm$    1          &      18    $\pm$   4            &      10   $\pm$    1         &     12   $\pm$   3   & 309     $\pm$   95   &  515    $\pm$   179     &  62  \\      
22187+5559V3 &$\le$0.6      &       $\le$ 5.6       &   $\le$0.6   &   $\le$6.3 & --&  --    &   69   \\ 
22187+5559V5 &$\le$0.7      &       $\le$ 6.7       &  $\le$0.8  &     $\le$7.2& --&  --     &  69   \\
22198+6336   &   5.4  $\pm$    0.7         &      12    $\pm$   3            &     6.3   $\pm$    0.9       &       16   $\pm$   3  & 287     $\pm$   83   &  251     $\pm$   79    &   64     \\               
22570+5912M2 &   2.1  $\pm$    0.3         &      3     $\pm$    1 $^{t}$        &     3.3   $\pm$    0.5       &       2.0   $\pm$ 0.9$^{t}$& 554     $\pm$   250   &  1305    $\pm$   650  &    79    \\               
23033+5951M1 &   4.3  $\pm$    0.6         &      9     $\pm$    2           &     9     $\pm$    1         &       14   $\pm$   3   & 349     $\pm$   104   &  469    $\pm$   127   &   73    \\           
23139+5939M1 &   2.3  $\pm$    0.4         &      6     $\pm$    2           &     10   $\pm$    1          &       116   $\pm$   3  & 301    $\pm$   145   &  491     $\pm$   130   &  79   \\             
G75-core      & 9      $\pm$  1            &       27    $\pm$   5            &     40   $\pm$    5          &       72   $\pm$   10 & 207     $\pm$   59   &  345     $\pm$   90      &   62   \\    
G08.14+0.22   &   14   $\pm$   2            &      15    $\pm$   4            &     27   $\pm$    3          &       36   $\pm$   7  & 395     $\pm$   158   &  318     $\pm$   102    &   42    \\       
G14.33-0.65   &   30   $\pm$   3        &         59     $\pm$  8             &     59   $\pm$    7         &       100   $\pm$   10 & 242    $\pm$   68   &  280     $\pm$   78     &  47   \\     
G14.99-0.67   &  8     $\pm$   2          &        15    $\pm$   4            &     10   $\pm$    2          &       11   $\pm$   4  & 257    $\pm$   105   &  439     $\pm$   200   &    48  \\    
G14.99-2      &   3.4  $\pm$    0.5         &      9     $\pm$    2           &     4.0   $\pm$    0.8       &       7     $\pm$   2 & 184     $\pm$   58   &  278     $\pm$   94     &   49  \\            
G15.02-0.62   &   5    $\pm$    0.9         &      6     $\pm$    1           &     8     $\pm$    1         &       12   $\pm$   2  & 406     $\pm$   141   &  325     $\pm$   98    &   49   \\            
G24.78+0.08   &   62   $\pm$   7           &       80    $\pm$     10         &     100     $\pm$    10      &       220   $\pm$   30& 197     $\pm$   81  &  116     $\pm$   47     &  25  \\             
G31.41+0.31  &     --    &       --       &    --  &     --& --&  --       &   33   \\
G34.82+0.35  &      6.2   $\pm$   0.9         &      9     $\pm$    3           &    7     $\pm$    1          &     10   $\pm$   2    & 316     $\pm$   130  &  322     $\pm$   104     &   46   \\          
G35.03+0.35  &      8     $\pm$   1          &       15    $\pm$   4            &    34   $\pm$    4           &     47   $\pm$   7    & 258     $\pm$   100   &  350     $\pm$   99     &  48  \\     
G35.20-0.74  &      29    $\pm$  3            &      46    $\pm$   8            &    60   $\pm$    7           &     90     $\pm$   10 & 330    $\pm$   95   &  349     $\pm$   94     &  52 \\         
G36.70+0.09  &  1.8$\pm$ 0.3        &     $\le$7        &   $\le$0.8  &   $\le$7.8 & $\ge$118 &  --      &  46 \\
G37.55+0.19  &  7$\pm$ 1       &       $\le$8.6       &    8$\pm$1  &  $\le$9.5 & $\ge$350 &  $\ge$362    &   43   \\ 
G40.28-0.22  &      7    $\pm$    1           &      10    $\pm$   3            &    18   $\pm$    2           &    18   $\pm$   5    & 312     $\pm$   124  &  446     $\pm$   165       &  45  \\    
G42.03+0.19  &  $\le$0.7      &       $\le$5.7       &   $\le$0.7 &     $\le$6.5& --&   --       &   55   \\
G42.30-0.30  &  2.8$\pm$ 0.4      &  $\le$5.6           &   2.4$\pm$0.4   &  $\le$6.8  &$\ge$273&  $\ge$193       &   55  \\
G42.70-0.15  &  $\le$0.7        &       $\le$6       &    $\le$0.7  &     $\le$7& --&  --       &   80  \\
G48.99-0.30  &     27   $\pm$   3            &       46    $\pm$   6            &    36   $\pm$    4          &      48   $\pm$   6   & 275     $\pm$   79   &  352     $\pm$   101      &   47   \\     
G49.41+0.33  &  $\le$1      &      $\le$8.4       &    $\le$0.9  &     $\le$9.4& --&  --       &  68  \\
G50.78+0.15  &  3.4$\pm$ 0.5        &    $\le$8.7         &    3.6$\pm$0.6  & $\le$9.3 &  $\ge$201&  $\ge$199     &  51   \\ 
G52.92+0.41  &      2.4   $\pm$   0.3         &      5     $\pm$    2           &     1.7$\pm$0.3            &   $\le$7 & 246    $\pm$   95   &  $\ge$124     &  51 \\ 
G53.04+0.11  &    3.3   $\pm$   0.5         &      4     $\pm$    1           &      7.8   $\pm$    1         &     8     $\pm$   2  & 496     $\pm$   196  &  586     $\pm$   186      &  60  \\             
G53.14+0.07  &    6.0   $\pm$   0.8         &      9     $\pm$    2           &      22   $\pm$    3         &     29   $\pm$   4   & 377     $\pm$   115   &  429     $\pm$   114    &  56   \\           
NGC7538-IRS1 &    7     $\pm$   1         &        12    $\pm$   4            &      32   $\pm$    4         &     55   $\pm$   9   & 318     $\pm$   137   &  317     $\pm$   88     &   55  \\      
NGC7538-IRS9 &    5.0   $\pm$   0.7         &      11    $\pm$   2            &      18   $\pm$    2         &     29   $\pm$   5   &322     $\pm$   93 &  439     $\pm$   119     &  71 \\ 
 \bottomrule
\end{longtable}
$^{t}$ tentative detection, as explained in Sect.~\ref{res}.



\appendix
\twocolumn
\section{Fit results}
In this appendix, the results of the fitting procedure to the HN$^{13}$C(1-0), H$^{15}$NC(1-0),  H$^{13}$CN(1-0) and HC$^{15}$N(1-0) lines of all sources are shown. The method is explained in Sect.~\ref{fitting}.

\onecolumn
\begin{longtable}{l*{5}{c}}
\caption{\label{fit1}Values obtained with the fitting procedure described in Sect.~\ref{res} to the HN$^{13}$C lines. In the second and in the third columns the centroid velocities and the FWHM are listed. In the fourth column the total column densities derived for the sources are listed, and in the fifth column there are the opacity of the lines. In the sixth column the r.m.s. of observed lines are listed. The cases in which the line is not detected, and therefore only column density upper limits could be obtained as explained in Sect.~\ref{res}, are indicated with -- .}
  \tabularnewline \hline \hline
  Source & \multicolumn{5}{c}{HN$^{13}$C(1-0)}   \\
  &$v_{\rm LSR}$ & $\Delta v_{\rm 1/2}$ & $N_{\rm tot}$ & $\tau$ & $\sigma$ \\
 &(km s$^{-1}$) & (km s$^{-1}$) & ($\times$10$^{12}$cm$^{-2}$) &    & (mK) \\
 \hline
\endfirsthead
\caption{Continued.} \\
\hline \hline
 Source & \multicolumn{5}{c}{HN$^{13}$C(1-0)}   \\
  &$v_{\rm LSR}$ & $\Delta v_{\rm 1/2}$ & $N_{\rm tot}$ & $\tau$ & $\sigma$ \\
 &(km s$^{-1}$) & (km s$^{-1}$) & ($\times$10$^{12}$cm$^{-2}$) &    & (mK) \\
 \hline
 \endhead
I00117-MM1     &       -36.03$\pm$4        &     1.6$\pm$0.1    &   1.5$\pm$ 0.3    &   0.0051$\pm$0.0006 & 16 \\
I04579-VLA1   &           --   &    --    &   $\le$1.1    &  --&  27\\
18089-1732M1 &         33.02$\pm$0.08     &    3.8$\pm$0.01    &   35    $\pm$  4          &   0.0087$\pm$0.0004  & 11\\
18089-1732M4  &       33.20$\pm$0.06     &    1.58$\pm$0.08    &     3.5  $\pm$    0.5     &   0.0120$\pm$0.0008 &  14 \\
18151-1208M1  &       33.31$\pm$0.04     &    2.01$\pm$0.05    &     8    $\pm$    1       &   0.0211$\pm$0.0007 & 15\\
18151-1208M2  &        29.78$\pm$0.07    &    2.51$\pm$0.07    &      8    $\pm$    1      &  0.0126$\pm$0.0005 & 15\\
18182-1433M1  &        59.86$\pm$0.06     &    3.42$\pm$0.07    &      16   $\pm$   2      &    0.0123$\pm$0.0004  & 9 \\
18264-1152M1  &        43.68$\pm$0.05     &    2.76$\pm$0.05    &      15   $\pm$   2       &   0.0156$\pm$0.0004 & 10 \\
18272-1217M1  &        34.0$\pm$0.1     &    1.6$\pm$0.2    &     1.4$\pm$ 0.3 &  0.0023$\pm$0.0002  & 14 \\
18290-0924M2  &       --   &    --    &   $\le$0.6    &   -- & 15 \\
18306-0835M1  &        78.15$\pm$0.05     &    2.53$\pm$0.04    &      10    $\pm$  1       &   0.0188$\pm$0.0004  & 11\\
18306-0835M2  &        76.7$\pm$0.1    &    2.23$\pm$0.08    &    4.7$\pm$ 0.6   &  0.0112$\pm$0.0005 & 15 \\
18308-0841M1  &        76.78$\pm$0.04     &    2.40$\pm$0.05    &      9     $\pm$   1      &   0.0204$\pm$0.0006 & 16\\
18310-0825M2  &        84.92$\pm$0.08     &    3.15$\pm$0.09    &      9     $\pm$   1     &   0.0146$\pm$0.0006  & 15\\
18310-0825M3  &        85.71$\pm$0.09    &    2.9$\pm$0.1    &     6.0   $\pm$   0.8     &   0.0127$\pm$0.0006 & 15\\
18372-0541M1  &        22.3$\pm$0.2     &    3.3$\pm$0.3    &   3.5$\pm$ 0.7  &  0.0015$\pm$0.0002  &  18\\
18385-0512M1  &        25.5$\pm$0.1     &    1.6$\pm$0.3    &   1.7$\pm$ 0.4  &  0.0031$\pm$0.0007 & 19 \\
18445-0222M3  &       112.13$\pm$0.9    &    1.75$\pm$0.06    &     1.6   $\pm$  0.2      &    0.0270$\pm$0.0006  & 13 \\
18454-0136M1  &        38.7$\pm$0.2    &    2.2$\pm$0.2    &      2.9   $\pm$  0.5     &   0.0086$\pm$0.0008 & 13\\
18472-0022M1  &        49.7$\pm$0.1     &    3.0$\pm$0.1    &      4.5   $\pm$  0.6     &   0.0101$\pm$0.0006  & 15\\
18488+0000M1  &        83.39$\pm$0.05    &    3.04$\pm$0.08    &    6.5   $\pm$  0.8       &   0.0114$\pm$0.0004     & 8\\
18517+0437M1  &       44.02$\pm$0.03     &    2.89$\pm$0.04    &    7.8   $\pm$  0.9       &   0.0110$\pm$0.0002&  7\\
18521+0134M1  &        76.8$\pm$0.1     &    2.8$\pm$0.2    &      3.1   $\pm$  0.5     &   0.0055$\pm$0.0005  & 13\\
19012+0536M1  &        66.3$\pm$0.1     &    3.0$\pm$0.1    &      3.5   $\pm$  0.5     &   0.0058$\pm$0.0003 & 10 \\
19035+0641M1  &        32.7$\pm$0.1    &    4.2$\pm$0.2    &    6.6$\pm$ 0.9  &   0.0046$\pm$0.0003  & 15\\
19095+0930    &        44.5$\pm$0.2    &    5.1$\pm$0.2    &  9$\pm$ 1     &   0.0041$\pm$0.0003 & 17\\
19282+1814M1  &        23.0$\pm$0.1    &    1.30$\pm$0.06    &     4.0   $\pm$   0.6    &  0.0111$\pm$0.0007& 15 \\
19410+2336M1  &        22.62$\pm$0.08    &    1.98$\pm$0.04    &     7.2   $\pm$   0.9    &  0.0244$\pm$0.0007   &  8 \\
19411+2306M1  &        29.5$\pm$0.2     &    2.2$\pm$0.1    &     1.9$\pm$ 0.3 &  0.0080$\pm$0.0005   & 18 \\
19413+2332M1  &        20.7$\pm$0.1     &    2.1$\pm$0.1    &     2.6$\pm$ 0.4 &  0.0065$\pm$0.0005   & 13  \\
ON1           &        11.83$\pm$0.06     &    3.6$\pm$0.1    &        24   $\pm$   3    &   0.0065$\pm$0.0004   & 15\\
20126+4104M1  &        -3.69$\pm$0.04    &    2.15$\pm$0.03    &      26   $\pm$   3     &   0.0172$\pm$0.0004  & 13 \\
20216+4107M1  &        -1.68$\pm$0.09     &    1.47$\pm$0.08    &      4.8  $\pm$    0.7    &  0.0080$\pm$0.0006& 17 \\
20319+3958M1  &       8.5$\pm$0.1     &    1.32$\pm$0.08    &       2.8  $\pm$    0.4   &   0.0045$\pm$0.0004& 12 \\
20332+4124M1  &       -2.8$\pm$0.2     &    1.9$\pm$0.2    &        1.6  $\pm$    0.3  &   0.0061$\pm$0.0008 & 15 \\
20343+4129M1  &       11.40$\pm$0.08     &    2.76$\pm$0.07    &      9    $\pm$    1      &  0.0083$\pm$0.0003 & 13\\
22187+5559V3  &      --   &    --    &    $\le$0.6   &   --  & 15\\
22187+5559V5  &       --   &    --    &    $\le$0.7   &  -- & 17 \\
22198+6336    &        -11.1$\pm$0.1    &    1.57$\pm$0.07    &       5.4  $\pm$    0.7     &   0.0059$\pm$0.0004& 11 \\
22570+5912M2  &       -47.74$\pm$0.08     &    1.8$\pm$0.1    &      2.1  $\pm$    0.3      &   0.0066$\pm$0.0007& 15 \\
23033+5951M1  &        -53.12$\pm$0.09    &    2.28$\pm$0.09    &    4.3  $\pm$    0.6        &  0.019$\pm$0.001 & 14 \\
23139+5939M1  &       -44.1$\pm$0.2    &    2.1$\pm$0.1    &      2.3  $\pm$    0.4      &   0.0049$\pm$0.0005& 15 \\
G75-core      &       0.3$\pm$0.1   &    3.8$\pm$0.1    &    9      $\pm$  1         &   0.0056$\pm$0.0003& 13 \\
G08.14+0.22   &       19.2$\pm$0.1    &    4.2$\pm$0.1    &       14   $\pm$   2        &   0.0089$\pm$0.0005& 18 \\
G14.33-0.65   &       22.53$\pm$0.03    &    3.12$\pm$0.05    &      30   $\pm$   3     &   0.0281$\pm$0.0006& 19  \\
G14.99-0.67  &       18.71$\pm$0.09     &    1.4$\pm$0.3    &     8     $\pm$   2       &   0.073$\pm$0.019 & 21 \\
G14.99-2     &       22.10$\pm$0.08    &    1.60$\pm$0.09    &      3.4  $\pm$    0.5      &   0.029$\pm$0.005  & 21\\
G15.02-0.62  &       19.3$\pm$0.2   &    3.9$\pm$0.3    &      5    $\pm$    0.9      &   0.042$\pm$0.005& 20 \\
G24.78+0.08  &       110.4$\pm$0.1     &    4.14$\pm$0.07    &      62   $\pm$   7        &   0.0123$\pm$0.0003  & 17 \\
G31.41+0.31  &       --     &    --    &   --     &  --   & 21\\
G34.82+0.35  &        56.9$\pm$0.1     &    2.3$\pm$0.1    &       6.2   $\pm$   0.9     &   0.015$\pm$0.001  & 18 \\
G35.03+0.35  &        53.7$\pm$0.1    &    3.5$\pm$0.1    &       8     $\pm$   1       &  0.0102$\pm$0.0005  & 18\\
G35.20-0.74  &       34.10$\pm$0.06    &    3.42$\pm$0.06    &       29    $\pm$  3        &  0.0237$\pm$0.0006& 19 \\
G36.70+0.09  &       51.9$\pm$0.4    &    1.25$\pm$0.07    &   1.8$\pm$ 0.3     &   0.0079$\pm$0.0006 & 20\\
G37.55+0.19  &       84.987$\pm$0.05    &    2.7$\pm$0.1    &     7$\pm$ 1  &    0.0096$\pm$0.0006  & 20\\
G40.28-0.22  &       73.0$\pm$0.1    &    3.7$\pm$0.2    &         7    $\pm$    1     &   0.0069$\pm$0.0006 & 19 \\
G42.03+0.19  &        --     &    --    &    $\le$0.7     &  --  & 16\\
G42.30-0.30  &        27.9$\pm$0.2   &    3.4$\pm$0.2    &  2.8$\pm$ 0.4    &   0.0046$\pm$0.0004 & 15 \\
G42.70-0.15 &       --     &    --    &  $\le$0.7     &    --  & 17\\
G48.99-0.30 &        67.93$\pm$0.04    &    2.84$\pm$0.05    &      27   $\pm$   3         &  0.049$\pm$0.001 & 15\\
G49.41+0.33 &      --  &    --    &    $\le$1    &   -- & 24\\
G50.78+0.15 &        42.36$\pm$0.09     &    2.2$\pm$0.1    &  3.4$\pm$ 0.5      &   0.0080$\pm$0.0005 & 23 \\
G52.92+0.41 &        45.5$\pm$0.1     &    1.7$\pm$0.1    &      2.4   $\pm$   0.3      &    0.0078$\pm$0.0006 & 15 \\
G53.04+0.11 &       4.7$\pm$0.2     &    3.0$\pm$0.1    &       3.3   $\pm$   0.5     &   0.0073$\pm$0.0005  & 15\\
G53.14+0.07 &        21.71$\pm$0.08    &    2.26$\pm$0.07    &      6.0   $\pm$   0.8      &   0.0106$\pm$0.0005 & 9\\
NGC7538-IRS1 &       -57.57$\pm$0.05    &    3.0$\pm$0.3    &       7     $\pm$   1      &   0.008$\pm$0.001  & 15 \\
NGC7538-IRS9 &       -57.03$\pm$0.08     &    2.4$\pm$0.1    &      5.0   $\pm$   0.7      &   0.0094$\pm$0.0007 & 12 \\
\bottomrule
 \end{longtable}
  $^{t}$ tentative detection, as explained in Sect.~\ref{res}.
\newpage
\begin{longtable}{l*{5}{c}}
\caption{\label{fit2} Same as Table \ref{fit1} for the H$^{15}$NC(1--0) lines.}
  \tabularnewline \hline \hline
Source & \multicolumn{5}{c}{H$^{15}$NC(1-0)}   \\
  &$v_{\rm LSR}$ & $\Delta v_{\rm 1/2}$ & $N_{\rm tot}$ & $\tau$ & $\sigma$ \\
 &(km s$^{-1}$) & (km s$^{-1}$) & ($\times$10$^{12}$cm$^{-2}$) &    & (mK) \\
 \hline
\endfirsthead
\caption{Continued.} \\
\hline \hline
 Source & \multicolumn{5}{c}{H$^{15}$NC(1-0)}   \\
  &$v_{\rm LSR}$ & $\Delta v_{\rm 1/2}$ & $N_{\rm tot}$ & $\tau$ & $\sigma$ \\
 &(km s$^{-1}$) & (km s$^{-1}$) & ($\times$10$^{12}$cm$^{-2}$) &    & (mK) \\
 \hline
 \endhead
I00117-MM1    &     --  &  --      &  $\le$5.9   &   -- &  16  \\
I04579-VLA1  &   --   &  --           &   $\le$ 10.1   &   -- &   28 \\
18089-1732M1 &    32.3$\pm$0.2    &  3.5$\pm$0.3     &  47    $\pm$   8        &    0.0010$\pm$0.0001&   10 \\
18089-1732M4 &     33.7$\pm$0.4    &  1.6$\pm$0.3     &    6    $\pm$     2      &   0.0022$\pm$0.0006&  14  \\
18151-1208M1  &     33.0$\pm$0.3    &  2.0$\pm$0.2     &  15   $\pm$    3         &  0.0040$\pm$0.0005 &  15  \\
18151-1208M2 &     29.4$\pm$0.1   &  2.5$\pm$0.4     &     13   $\pm$    3      &    0.0021$\pm$0.0005 &  16  \\
18182-1433M1 &     59.6$\pm$0.3    &  3.7$\pm$0.2     &    25   $\pm$    4       &   0.0018$\pm$0.0002&   9 \\
18264-1152M1  &     43.9$\pm$0.1   &  2.5$\pm$0.1     &   29   $\pm$    4        &  0.0034$\pm$0.0002 &   10 \\
18272-1217M1 &    --   &  --     &    $\le$7.1  &  -- &  14  \\
18290-0924M2 &    --  &  --     &   $\le$6.2   &   --&   17 \\
18306-0835M1  &    78.1$\pm$0.3    &  2.2$\pm$0.3     &     10    $\pm$   2   &   0.0023$\pm$0.0005 &   12 \\
18306-0835M2  &    &  --     &    $\le$4.4      &    --&   16 \\
18308-0841M1  &     76.6$\pm$0.2   &  2.0$\pm$0.3     &   10    $\pm$   2        &   0.0029$\pm$0.0006&  16   \\
18310-0825M2 &    86.3$\pm$0.2   &  2.7$\pm$0.4     &    10    $\pm$   1       &   0.0022$\pm$0.0005&   16 \\
18310-0825M3 &    85.5$\pm$0.1   &  1.7$\pm$0.3     &     7     $\pm$    2     &    0.0025$\pm$0.0007&  15  \\
18372-0541M1  &    --   &  --     &   $\le$1.2  &    -- &   18 \\
18385-0512M1  &    --   &  --     &    $\le$9.1  &   -- &   19 \\
18445-0222M3 &    112.5$\pm$0.1    &  0.98$\pm$0.66     &    2.0   $\pm$   0.9     &   0.0045$\pm$0.0003&  14  \\
18454-0136M1  &    39.4$\pm$0.2   &  1.9$\pm$0.5     &   5     $\pm$   1        &  0.0017$\pm$0.0005 &  15  \\
18472-0022M1 &    49.1$\pm$0.3   &  1.8$\pm$0.5     &    5     $\pm$   2       &    0.0020$\pm$0.0008&  16  \\
18488+0000M1 &     84.7$\pm$0.6   &  2.5$\pm$0.3     &    7     $\pm$   1       &   0.0015$\pm$0.0003&  8  \\
18517+0437M1 &    43.9$\pm$0.4   &  2.6$\pm$0.1     &   13    $\pm$  2         &   0.0023$\pm$0.0002&   7 \\
18521+0134M1 &    75.9$\pm$0.5    &  2.1$\pm$0.7     &    5     $\pm$   2       &  0.0014$\pm$0.0006  &   13 \\
19012+0536M1 &    66.0$\pm$0.5    &  2.6$\pm$0.6     &    5     $\pm$   1$^{t}$       &   0.0010$\pm$0.0003 &   10 \\
19035+0641M1 &   --   &  --     &      $\le$7.4   &  --&  15  \\
19095+0930   &    --   &  --     &  $\le$9.4   &   -- &   17 \\
19282+1814M1  &     22.4$\pm$0.3   &  0.9$\pm$0.2     &    7     $\pm$    2  &   0.0032$\pm$0.0009  &   16 \\
19410+2336M1 &    22.8$\pm$0.2   &  1.71$\pm$0.07     &  14    $\pm$   2             &  0.0059$\pm$0.0003&   8 \\
19411+2306M1 &    --   &  --     &   $\le$5.3  &    --&   18 \\
19413+2332M1 &    --   &  --     &    $\le$5.1  &   --&   14 \\
ON1          &    11.2$\pm$0.1    &  3.2$\pm$0.5     &   40    $\pm$   9    &  0.0013$\pm$0.0001 &   15 \\
20126+4104M1 &    -3.6$\pm$0.1    &  1.89$\pm$0.08     &   51    $\pm$   7    &   0.0040$\pm$0.0002&   13 \\
20216+4107M1 &   -1,4$\pm$0.4   &  2.0$\pm$0.4     &   10    $\pm$     3  &   0.0014$\pm$0.0004 &   17 \\
20319+3958M1 &    8.9$\pm$0.3   &  1.8$\pm$0.4     &   7     $\pm$    2   &    0.0008$\pm$0.0003 &  12  \\
20332+4124M1 &    -3.2$\pm$0.2   &  1.2$\pm$0.5     &   3     $\pm$    1$^{t}$   &    0.0017$\pm$0.0009 &  16  \\
20343+4129M1 &    11.0$\pm$0.4   &  2.9$\pm$0.04     &  18    $\pm$   4     &   0.0017$\pm$0.0003 &   13 \\
22187+5559V3 &    --  &  --     &   $\le$ 5.6   &    --&   15 \\
22187+5559V5 &     --  &  --     &    $\le$ 6.7   &   --&  18  \\
22198+6336   &    -11.4$\pm$0.4    &  1.1$\pm$0.1     &      12    $\pm$   3     &   0.0020$\pm$0.0004&  11  \\
22570+5912M2 &    -48.0$\pm$0.3    &  0.8$\pm$0.3     &       3     $\pm$    1$^{t}$   &   0.0014$\pm$0.0005&   15 \\
23033+5951M1 &    -53.2$\pm$0.2   &  1.9$\pm$0.2     &     9     $\pm$    2     &   0.0050$\pm$0.0009 &   16 \\
23139+5939M1  &    -45.0$\pm$0.4   &  1.3$\pm$0.3     &       6     $\pm$    2   &    0.0023$\pm$0.0009&  15  \\
G75-core     &    0.1$\pm$0.4    &  3.7$\pm$0.3     &      27    $\pm$   5     &   0.0017$\pm$0.0002 &    13\\
G08.14+0.22  &     19.4$\pm$0.2    &  2.8$\pm$0.6     &     15    $\pm$   4      &    0.0014$\pm$0.0004 &  19  \\
G14.33-0.65  &    22.72$\pm$0.09    &  3.2$\pm$0.1     &    59     $\pm$  8       &    0.0057$\pm$0.0004&  20  \\
G14.99-0.67  &    18.7$\pm$0.1    &  1.2$\pm$0.3     &      15    $\pm$   4     &   0.018$\pm$0.005 &   21 \\
G14.99-2     &     22.4$\pm$0.1  &  2.0$\pm$0.2     &      9     $\pm$    2    &   0.006$\pm$0.001&   21 \\
G15.02-0.62  &    19.7$\pm$0.1   &  3.0$\pm$0.4     &      6     $\pm$    1    &    0.007$\pm$0.001&  21  \\
G24.78+0.08  &    110.1$\pm$0.3    &  3.5$\pm$0.2     &     80    $\pm$     10  &   0.0020$\pm$0.0002 &   22 \\
G31.41+0.31  &  --    & --       &   --  &   --&   23 \\
G34.82+0.35  &    57.1$\pm$0.3  &  2.4$\pm$0.6     &   9     $\pm$    3  &  0.0023$\pm$0.0008 &   19 \\
G35.03+0.35  &    52.9$\pm$0.3   &  3.1$\pm$0.7     &   15    $\pm$   4   &   0.0022$\pm$0.0007 &  19  \\
G35.20-0.74    &    34.44$\pm$0.08   &  3.0$\pm$0.3     &  46    $\pm$   8    &  0.0045$\pm$0.0006&  19  \\
G36.70+0.09   &   --   &  --     &   $\le$7    &   -- &   19 \\
G37.55+0.19  &    --   &  --     &    $\le$8.6  &    --&  20  \\
G40.28-0.22  &     74.0$\pm$0.3    &  5$\pm$1     &   10    $\pm$   3      &   0.0019$\pm$0.0006 &  20  \\
G42.03+0.19  &     --   &  --     &    $\le$5.7  &   -- &   16 \\
G42.30-0.30  &    --   &  --     &    $\le$5.6   &   -- &   16 \\
G42.70-0.15  &    --   &  --     &  $\le$6   &    --&   17 \\
G48.99-0.30  &     67.85$\pm$0.08    &  2.4$\pm$0.1     &   46    $\pm$   6   &   0.0101$\pm$0.0007&  14  \\
G49.41+0.33  &    --   &  --     &    $\le$8.4 &   -- &   23 \\
G50.78+0.15  &    --   &  --     &  $\le$8.7   &  --  &   24 \\
G52.92+0.41  &     46.0$\pm$0.1    &  1.7$\pm$0.6     &    5     $\pm$    2    &   0.0016$\pm$0.0003&   16 \\
G53.04+0.11  &    6.0$\pm$0.3   &  2.5     &   4     $\pm$    1     &   0.0010$\pm$0.0003  &  15  \\
G53.14+0.07  &    21.9$\pm$0.2  &  1.9$\pm$0.2     &    9     $\pm$    2    &   0.0020$\pm$0.0003 &   10 \\
NGC7538-IRS1 &    -58.4$\pm$0.3   &  2.3$\pm$0.8     &   12    $\pm$   4     &  0.0021$\pm$0.0009&   17 \\
NGC7538-IRS9  &     -56.5$\pm$0.2    &  1.7$\pm$0.2     &   11    $\pm$   2      &   0.0031$\pm$0.0005 &   12 \\
\bottomrule
 \end{longtable}
 $^{t}$ tentative detection, as explained in Sect.~\ref{res}.

\newpage

\begin{longtable}{l*{5}{c}}
\caption{\label{fit3}   Same as Table \ref{fit1} for the H$^{13}$CN lines. Note that in some cases (e.g. 18310-0825M3) for the two isotopologues the fitted line width can be different and this can be explained by a higher r.m.s. in a spectrum than in the other cases. The higher r.m.s. noise worsen the S/N ratio for the $^{15}$N-isotopologues which have weaker emission lines with respect to $^{13}$C-isotopologues.}
\tabularnewline \hline \hline
  Source & \multicolumn{5}{c}{H$^{13}$CN(1-0)}   \\
  &$v_{\rm LSR}$ & $\Delta v_{\rm 1/2}$ & $N_{\rm tot}$ & $\tau^{1}$ & $\sigma$ \\
 &(km s$^{-1}$) & (km s$^{-1}$) & ($\times$10$^{12}$cm$^{-2}$) &    & (mK) \\
 \hline
\endfirsthead
\caption{Continued.} \\
\hline \hline
 Source & \multicolumn{5}{c}{H$^{13}$CN(1-0)}   \\
  &$v_{\rm LSR}$ & $\Delta v_{\rm 1/2}$ & $N_{\rm tot}$ & $\tau^{1}$ & $\sigma$ \\
 &(km s$^{-1}$) & (km s$^{-1}$) & ($\times$10$^{12}$cm$^{-2}$) &    & (mK) \\
 \hline 
 \endhead
I00117-MM1    &     -36.2$\pm$0.2  &    2.1$\pm$ 0.2  &   2.7    $\pm$   0.4  &    0.0046$\pm$0.0005 &   15     \\   
I04579-VLA1    &     -16.7$\pm$0.2&    2.5$\pm$ 0.2  &  4.2$\pm$0.8   &    0.0061$\pm$0.0008 &   27    \\ 
18089-1732M1   &     32.95$\pm$0.07  &    4.0$\pm$ 0.1  &    78   $\pm$    9    &    0.0125$\pm$0.0004 &   11     \\  
18089-1732M4   &     33.4$\pm$0.1  &    2.9$\pm$ 0.1  &    7.0   $\pm$    1.0   &    0.0085$\pm$0.0005 &   15    \\  
18151-1208M1   &     33.41$\pm$0.04  &    2.30$\pm$ 0.05  &   14   $\pm$    2     &    0.0211$\pm$0.0006&   15    \\  
18151-1208M2   &     29.79$\pm$0.08  &    3.6$\pm$ 0.1  &   14   $\pm$    2   &    0.0108$\pm$0.0006 &   16    \\  
18182-1433M1   &    59.53$\pm$0.07  &    4.07$\pm$ 0.07  &    40   $\pm$    4     &     0.0169$\pm$0.0004  &  9   \\  
18264-1152M1   &    43.75$\pm$0.06 &    3.58$\pm$ 0.07  &    36   $\pm$    4  &    0.0193$\pm$0.0005 &   9   \\  
18272-1217M1   &    34.2$\pm$0.4  &    2.4$\pm$ 0.2  &    4.5$\pm$0.7 &    0.0033$\pm$0.0003 &   15    \\  
18290-0924M2   &   -- &    --  &     $\le$1&    --&   16   \\  
18306-0835M1   &     78.11$\pm$0.06 &    3.20$\pm$ 0.09  &    21   $\pm$    3    &     0.0206$\pm$0.0008&   13    \\  
18306-0835M2   &    76.7$\pm$0.2 &    3.5$\pm$ 0.1  &    6.9   $\pm$    0.9   &    0.0070$\pm$0.0004 &   15   \\  
18308-0841M1   &     77.0$\pm$0.1  &    3.1$\pm$ 0.1  &   12   $\pm$    2     &    0.0139$\pm$0.0006 &   16   \\  
18310-0825M2   &    84.5$\pm$0.2  &    4.8$\pm$ 0.2  &    17   $\pm$    2   &    0.0110$\pm$0.0005 &   15   \\  
18310-0825M3   &    85.4$\pm$0.3 &    3.8$\pm$ 0.1  &    5.9   $\pm$    0.8         &     0.0061$\pm$0.0004&   15   \\  
18372-0541M1   &     23.6$\pm$0.2  &    3.5$\pm$ 0.2  &    12$\pm$1  &    0.0032$\pm$0.0002 &   18   \\  
18385-0512M1   &     26.1$\pm$0.4  &    4.5$\pm$ 0.3  &    8   $\pm$    1  &     0.0035$\pm$0.0003&   19    \\  
18445-0222M3   &    112.0$\pm$0.2  &    3.0$\pm$ 0.2  &   1.9   $\pm$    0.3  &    0.0118$\pm$0.0008&   14    \\  
18454-0136M1   &     38.7$\pm$0.2  &    2.7$\pm$ 0.1  &    2.0   $\pm$    0.5   &    0.0097$\pm$0.0007&   13    \\  
18472-0022M1   &    49.8$\pm$0.2  &    3.8$\pm$ 0.2  &     5.1   $\pm$    0.7   &    0.0060$\pm$0.0004&   16    \\  
18488+0000M1   &    83.0$\pm$0.1  &    4.2$\pm$ 0.1  &     14   $\pm$    2      &    0.0116$\pm$0.0004&   8    \\  
18517+0437M1   &    43.93$\pm$0.05  &    3.25$\pm$ 0.02  &    23   $\pm$    2   &    0.0194$\pm$0.0002&   7    \\  
18521+0134M1   &    76.1$\pm$0.3 &    3.1$\pm$ 0.1  &    4.8$\pm$0.7 &     0.0051$\pm$0.0003 &   13   \\  
19012+0536M1   &    65.8$\pm$0.2 &    3.76$\pm$ 0.09  &    11   $\pm$    1  &    0.0101$\pm$0.0003&   9    \\  
19035+0641M1   &     33.1$\pm$0.4 &    5.0$\pm$ 0.3  &    9     $\pm$    1  &    0.0036$\pm$0.0003&   15   \\  
19095+0930     &      43.9$\pm$0.2  &    6.2$\pm$ 0.2  &    35   $\pm$    4 &      0.0083$\pm$0.0003 &   17   \\ 
19282+1814M1   &     23.0$\pm$0.1  &    1.36$\pm$ 0.07  &    3.8   $\pm$    0.6 &    0.0068$\pm$0.0005 &   16    \\  
19410+2336M1   &     22.57$\pm$0.04  &    2.35$\pm$ 0.05  &    14   $\pm$    2    &   0.0274$\pm$0.0007 &  8   \\  
19411+2306M1   &     29.3$\pm$0.1  &    2.0$\pm$ 0.2  &     2.6   $\pm$    0.5 &     0.0083$\pm$0.0009 &   18   \\  
19413+2332M1   &     20.2$\pm$0.1  &    2.12$\pm$ 0.07  &    6.5   $\pm$    0.8  &    0.0109$\pm$0.0005 &   13   \\  
ON1            &    11.9$\pm$0.1  &    3.82$\pm$ 0.09  &       35   $\pm$    4       &     0.0061$\pm$0.0002 &   15    \\  
20126+4104M1   &    -3.63$\pm$0.08 &    2.90$\pm$ 0.09  &     31   $\pm$    4        &     0.0101$\pm$0.0004 &   12    \\  
20216+4107M1   &    -1.7$\pm$0.1  &    1.8$\pm$ 0.1  &       4.5   $\pm$    0.7    &     0.0041$\pm$0.0004 &   17   \\  
20319+3958M1   &    8.20$\pm$0.07  &    1.5$\pm$ 0.1  &    3.2$\pm$0.5  &    0.0030$\pm$0.0003  &   12   \\  
20332+4124M1   &    -2.7$\pm$0.1 &    2.8$\pm$ 0.2  &    3.4   $\pm$    0.5 &    0.0062$\pm$0.0006 &   16   \\
20343+4129M1   &     11.52$\pm$0.07 &    2.59$\pm$ 0.08  &    10   $\pm$    1    &    0.0069$\pm$0.0003 &   13    \\  
22187+5559V3   &    -- &    --  &    $\le$0.6  &    --&   15   \\  
22187+5559V5   &     -- &    --  &    $\le$0.8 &     -- &   18   \\   
22198+6336    &     -11.3$\pm$0.2  &    1.84$\pm$ 0.08  &     6.3   $\pm$    0.9   &     0.0040$\pm$0.0002&   11    \\ 
22570+5912M2   &    -48.0$\pm$0.3  &    2.5$\pm$ 0.2  &     3.3   $\pm$    0.5   &     0.0050$\pm$0.0005&   15    \\  
23033+5951M1   &    -53.16$\pm$0.05 &    2.65$\pm$ 0.06  &     9     $\pm$    1     &    0.0222$\pm$0.0007&   15   \\  
23139+5939M1   &    -44.51$\pm$0.08  &    3.0$\pm$ 0.1  &     10   $\pm$    1      &     0.0100$\pm$0.0005&   15   \\  
G75-core      &     0.1$\pm$0.08 &    3.99$\pm$ 0.09  &    40   $\pm$    5       &     0.0150$\pm$0.0005 &   13   \\ 
G08.14+0.22   &      19.07$\pm$0.06  &    4.26$\pm$ 0.07  &     27   $\pm$    3      &     0.0112$\pm$0.0003&   19    \\ 
G14.33-0.65   &     22.41$\pm$0.07  &    3.6$\pm$ 0.1  &     59   $\pm$    7      &    0.0318$\pm$0.0012 &   19   \\ 
G14.99-0.67   &     18.71$\pm$0.04  &    1.8$\pm$ 0.2  &    10   $\pm$    2       &    0.049$\pm$0.008&   21   \\ 
G14.99-2      &     21.86$\pm$0.09  &    1.4$\pm$ 0.2  &    4.0   $\pm$    0.8    &    0.026$\pm$0.004 &   21    \\ 
G15.02-0.62   &      18.91$\pm$0.08  &    3.1$\pm$ 0.1  &    8     $\pm$    1      &    0.055$\pm$0.003&   19    \\ 
G24.78+0.08   &     111.2$\pm$0.2 &    4  &    100     $\pm$    10   &     0.0148$\pm$0.0008 &   17   \\ 
G31.41+0.31   &     -- &    --  &    -- &    --&   19   \\ 
G34.82+0.35   &     56.9$\pm$0.1  &    2.5$\pm$ 0.1  &   7     $\pm$    1      &    0.0109$\pm$0.0007 &   18    \\ 
G35.03+0.35   &      53.31$\pm$0.06 &    4.04$\pm$ 0.06  &    34   $\pm$    4      &    0.0245$\pm$0.0005&   17    \\ 
G35.20-0.74   &      34.57$\pm$0.07 &    5.1$\pm$ 0.1  &    60   $\pm$    7      &    0.0223$\pm$0.0007 &   18    \\ 
G36.70+0.09   &     -- &    --  &   $\leq$0.8 &    -- &   19   \\ 
G37.55+0.19   &      84.2$\pm$0.5  &    5.0$\pm$ 0.4  &    8$\pm$1 &     0.0042$\pm$0.0004&   18    \\ 
G40.28-0.22   &      73.5$\pm$0.3  &    8.7$\pm$ 0.7  &    18   $\pm$    2  &     0.0052$\pm$0.0005&   19     \\ 
G42.03+0.19   &      -- &    --  &    $\le$0.7 &    --&   16    \\ 
G42.30-0.30   &      27.3$\pm$0.3  &    4.3$\pm$ 0.4  &    2.4$\pm$0.4 &    0.0021$\pm$0.0003&   16    \\ 
G42.70-0.15   &      --  &    --  &   $\le$0.7  &    --&   16    \\ 
G48.99-0.30   &      67.89$\pm$0.07  &    3.66$\pm$ 0.08  &   36   $\pm$    4   &    0.034$\pm$0.001 &   14    \\ 
G49.41+0.33   &      --  &    --  &   $\le$0.9  &    --&   22    \\ 
G50.78+0.15   &      42.3$\pm$0.2  &    2.4$\pm$ 0.2  &   3.6$\pm$0.6  &     0.0056$\pm$0.0007&   21   \\ 
G52.92+0.41   &      45.9$\pm$0.3  &    1.8$\pm$ 0.2  &   1.7$\pm$0.3   &    0.0035$\pm$0.0005&   15     \\ 
G53.04+0.11   &      5.0$\pm$0.2 &    3.30$\pm$ 0.09  &   7.8   $\pm$    1    &    0.0104$\pm$0.0004&   15   \\ 
G53.14+0.07   &     21.59$\pm$0.09  &    3.37$\pm$ 0.08  &    22   $\pm$    3   &    0.0173$\pm$0.0006 &   9    \\ 
NGC7538-IRS1   &     -57.47$\pm$0.05  &    3.28$\pm$ 0.06  &   32   $\pm$    4    &    0.0244$\pm$0.0006&   15    \\ 
NGC7538-IRS9   &     -57.4$\pm$0.1 &    3.9$\pm$ 0.1  &    18   $\pm$    2   &     0.0140$\pm$0.0005 &   12   \\ 
\bottomrule
 \end{longtable}
$^{1}$ opacity of the main hyperfine component;\\
$^{t}$ tentative detection, as explained in Sect.~\ref{res}.
\newpage
\begin{longtable}{l*{5}{c}}
\caption{\label{fit4}  Same as Table \ref{fit1} for the HC$^{15}$N(1-0) lines. }
\tabularnewline \hline \hline
   Source & \multicolumn{5}{c}{HC$^{15}$N(1-0)}   \\
  &$v_{\rm LSR}$ & $\Delta v_{\rm 1/2}$ & $N_{\rm tot}$ & $\tau$ & $\sigma$ \\
 &(km s$^{-1}$) & (km s$^{-1}$) & ($\times$10$^{12}$cm$^{-2}$) &    & (mK) \\
 \hline
\endfirsthead
\caption{Continued.} \\
\hline \hline
 Source & \multicolumn{5}{c}{HC$^{15}$N(1-0)}   \\
  &$v_{\rm LSR}$ & $\Delta v_{\rm 1/2}$ & $N_{\rm tot}$ & $\tau$ & $\sigma$ \\
 &(km s$^{-1}$) & (km s$^{-1}$) & ($\times$10$^{12}$cm$^{-2}$) &    & (mK) \\
 \hline 
 \endhead
I00117-MM1   &   -36.1$\pm$0.4   &   2.0$\pm$ 0.7  &  5      $\pm$  2 $^{t}$  &  0.0015$\pm$0.0008   &  17\\
I04579-VLA1  &   -- &   --  &   $\le$11  &  --    &  29\\
18089-1732M1 &    33.1$\pm$0.1   &   4.4$\pm$ 0.1  &   140   $\pm$   20   &  0.0035$\pm$0.0001   &  12\\
18089-1732M4 &    33.7$\pm$0.5  &   2.6$\pm$ 0.5  &  7     $\pm$   2  &   0.0018$\pm$0.0005   &  17\\
18151-1208M1  &   33.5$\pm$0.1 &   2.6$\pm$ 0.2  &  23   $\pm$   4   &   0.0052$\pm$0.0007   &  17\\
18151-1208M2  &    30.1$\pm$0.3  &   3.8$\pm$ 0.6  &  18   $\pm$   4   &  0.0023$\pm$0.0005   &  17\\
18182-1433M1 &   59.8$\pm$0.2   &   4.0$\pm$ 0.2  & 48   $\pm$   6   &  0.0037$\pm$0.0002   &  10\\
18264-1152M1  &   43.8$\pm$0.1  &   3.1$\pm$ 0.2  & 48   $\pm$   7    &  0.0052$\pm$0.0004   &  12\\
18272-1217M1  &   --&   --  &   $\le$9.3 &  --   &  16\\
18290-0924M2 &   --&   --  &  $\leq$7.1   &   --   &  18\\
18306-0835M1 &   78.1$\pm$0.2   &   2.3$\pm$ 0.2  &    17   $\pm$   3   &  0.0040$\pm$0.0006    &  13\\
18306-0835M2  &    77.0$\pm$0.1 &   2.0$\pm$ 0.8  & 4     $\pm$   2    &  0.0012$\pm$0.0007   &  15\\
18308-0841M1 &    76.7$\pm$0.3   &   2.5$\pm$ 0.4  &   11   $\pm$   3   &  0.0028$\pm$0.0006   &  17\\
18310-0825M2 &   84.2$\pm$0.4   &   4.8$\pm$ 0.6  & 19   $\pm$   4     &   0.0023$\pm$0.0004   &  15\\
18310-0825M3 &    85.2$\pm$0.1  &   4$\pm$ 1  &    8     $\pm$   3 $^{t}$ &  0.0014$\pm$0.0005   &  16\\
18372-0541M1  &    --  &   --  &  $\le$13  &   --   &  18\\
18385-0512M1 &    23.4$\pm$0.2  &   5$\pm$ 1  & 19   $\pm$   5  &  0.0012$\pm$0.0003   &  19\\
18445-0222M3 &   113.6$\pm$0.6   &   1.5$\pm$ 0.5  &    1.6   $\pm$   0.6  &   0.0040$\pm$0.0006   &  16\\
18454-0136M1 &   37.9$\pm$0.2   &   2.8$\pm$ 0.5  &  6     $\pm$   2  &   0.0017$\pm$0.0005   &  16\\
18472-0022M1 &    48.9$\pm$0.9  &   1.0$\pm$ 0.5  &  3     $\pm$   1  &  0.0030$\pm$0.0009   &  19\\
18488+0000M1 &   83.1$\pm$0.3   &   4.1$\pm$ 0.3  & 15   $\pm$   2    &   0.0022$\pm$0.0003   &  9\\
18517+0437M1 &    43.1$\pm$0.3  &   3.47$\pm$ 0.08  &  35   $\pm$   4   &   0.0051$\pm$0.0002   &  9\\
18521+0134M1 &   --  &   --  & $\le$6.3 &  --    &  15\\
19012+0536M1 &   65.6$\pm$0.6  &   3.0$\pm$ 0.3  &   12   $\pm$   2   &  0.0024$\pm$0.0004   &  11\\
19035+0641M1 &   31.4$\pm$0.3   &   5$\pm$ 1  &   14   $\pm$   4 $^{t}$   &  0.0010$\pm$0.0003    &  15\\
19095+0930   &   44.0$\pm$0.4  &   5.9$\pm$ 0.6  & 46   $\pm$   8   &   0.0020$\pm$0.003   &  17\\
19282+1814M1 &   24.8$\pm$0.2   &   0.86  & 4     $\pm$   2    &   0.0017$\pm$0.0005   &  19\\
19410+2336M1  &   22.3$\pm$0.2 &   1.98$\pm$ 0.08  &  19   $\pm$   3    &  0.0077$\pm$0.0004   &  8\\
19411+2306M1  &   29.0$\pm$0.3 &   1.3$\pm$ 0.4  &  3     $\pm$   1    &  0.003$\pm$0.001   &  21\\
19413+2332M1 &   20.4$\pm$0.2  &   2.0$\pm$ 0.4  &  6     $\pm$   2    &   0.0019$\pm$0.0006   &  15\\
ON1          &   11.3$\pm$0.3  &   3.8$\pm$ 0.3  & 55   $\pm$   9   &  0.0017$\pm$0.0004   &  18\\
20126+4104M1 &   -3.8$\pm$0.4  &   3.5$\pm$ 0.2  &  59   $\pm$   9  &  0.0028$\pm$0.0003   &  13\\
20216+4107M1 &   -2.0$\pm$0.2  &   2.0$\pm$ 0.6  & 9     $\pm$   3  &   0.0013$\pm$0.0005   &  20\\
20319+3958M1 &   --   &   --  & $\le$8.2   &  --    &  13\\
20332+4124M1 &    -2.$\pm$1   &   1.9$\pm$ 0.6  &  4     $\pm$   1 &   0.0020$\pm$0.0008   &  15\\
20343+4129M1 &   11.8$\pm$0.2  &   2.0$\pm$ 0.3  &  12   $\pm$   3  &  0.0018$\pm$0.0004   &  13\\
22187+5559V3 &    --  &   --  & $\le$6.3   & --   &  16\\
22187+5559V5 &   --  &   --  &   $\le$7.2& --   &  18\\
22198+6336   &   -11.6$\pm$0.2  &   2.1$\pm$ 0.3  &   16   $\pm$   3    &  0.0016$\pm$0.0003    &  12\\
22570+5912M2  &   -48.1$\pm$0.3   &   0.85  &  2.0   $\pm$   0.9 $^{t}$  &  0.0014$\pm$0.0004   &  15\\
23033+5951M1  &   -53.4$\pm$0.5   &   3.1$\pm$ 0.3  &  14   $\pm$   3     &   0.0055$\pm$0.0008   &  16\\
23139+5939M1 &   -44.1$\pm$0.3   &   2.4$\pm$ 0.2  &  116   $\pm$   3    &   0.0037$\pm$0.0005   &  15\\
G75-core     &    -0.05$\pm$0.3 &   4.4$\pm$ 0.2  &  72   $\pm$   10    &   0.0044$\pm$0.0003   &  14\\
G08.14+0.22  &   19.4$\pm$0.4  &   4.3$\pm$ 0.4  &  36   $\pm$   7     &  0.0027$\pm$0.0004   &  20\\
G14.33-0.65  &   22.17$\pm$0.09   &   3.6$\pm$ 0.1  & 100   $\pm$   10    &  0.0098$\pm$0.0005    &  19\\
G14.99-0.67   &    18.9$\pm$0.3   &   1.3$\pm$ 0.4  &  11   $\pm$   4     &  0.014$\pm$0.006   &  21\\
G14.99-2     &   22.10$\pm$0.07 &   1.5$\pm$ 0.2  &  7     $\pm$   2    & 0 0.008$\pm$0.002   &  21\\
G15.02-0.62  &   18.9$\pm$0.3   &   3.6$\pm$ 0.3  &  12   $\pm$   2     &  0.013$\pm$0.001   &  20\\
G24.78+0.08  &   110.9$\pm$0.2   &   6.0$\pm$ 0.2  &  220   $\pm$   30   &   0.0037$\pm$0.0002   &  20\\
G31.41+0.31  &  --   &  --   &  -- &  --    &  19\\
G34.82+0.35   &   56.3$\pm$0.3   &   3.1$\pm$ 0.5  &  10   $\pm$   2      &  0.0023$\pm$0.0005   &  19\\
G35.03+0.35  &   52.8$\pm$0.2 &   4.6$\pm$ 0.3  &  47   $\pm$   7      &  0.0053$\pm$0.0004   &  19\\
G35.20-0.74  &    34.2$\pm$0.2   &   5.2$\pm$ 0.2  &  90     $\pm$   10   &   0.0061$\pm$0.0004   &  20\\
G36.70+0.09  &   --  &   --  &    $\le$7.8&  --   &  20\\
G37.55+0.19  &  --  &   --  & $\le$9.5   & --   &  19\\
G40.28-0.22  &   73.6$\pm$0.4 &   3.4$\pm$ 0.7  &  18   $\pm$   5    &  0.0023$\pm$0.0006   &  20\\
G42.03+0.19  &   --  &   --  &   $\le$6.5 &  --   &  16\\
G42.30-0.30   &    --  &   --  &  $\le$6.8  &   --   &  17\\
G42.70-0.15  &   --   &   --  & $\le$7 &  --   &  18\\
G48.99-0.30  &   67.67$\pm$0.09   &   3.7$\pm$ 0.1  &   48   $\pm$   6   &  0.0079$\pm$0.0005    &  14\\
G49.41+0.33  &   --  &   --  &  $\le$9.4&  --   &  24\\
G50.78+0.15  &   -- &   --  &  $\le$9.3   &   --   &  23\\
G52.92+0.41  &   --  &   --  & $\le$7   &  --   &  18\\
G53.04+0.11   &   4.9$\pm$0.2  &   2.6$\pm$ 0.5  & 8     $\pm$   2  &  0.0023$\pm$0.0006   &  19\\
G53.14+0.07  &    22.0$\pm$0.1   &   3.5$\pm$ 0.2  & 29   $\pm$   4   &  0.0040$\pm$0.0003   &  11\\
NGC7538-IRS1 &   -57.54$\pm$0.07  &   3.3$\pm$ 0.2  & 55   $\pm$   9   &   0.0075$\pm$0.0008   &  16\\
NGC7538-IRS9 &   -57.4$\pm$0.3   &   3.2$\pm$ 0.2  & 29   $\pm$   5  &  0.0050$\pm$0.0005   &  12\\
\bottomrule
 \end{longtable}
$^{t}$ tentative detection, as explained in Sect.~\ref{res}.

\twocolumn
\section{Spectra}
\label{ap-spectra}
In this appendix, all spectra of HN$^{13}$C(1-0), H$^{13}$CN(1-0), H$^{15}$NC(1-0) and HC$^{15}$NC(1-0) transitions for all the sources are shown.

\begin{figure*}
\centering
\includegraphics[width=30pc]{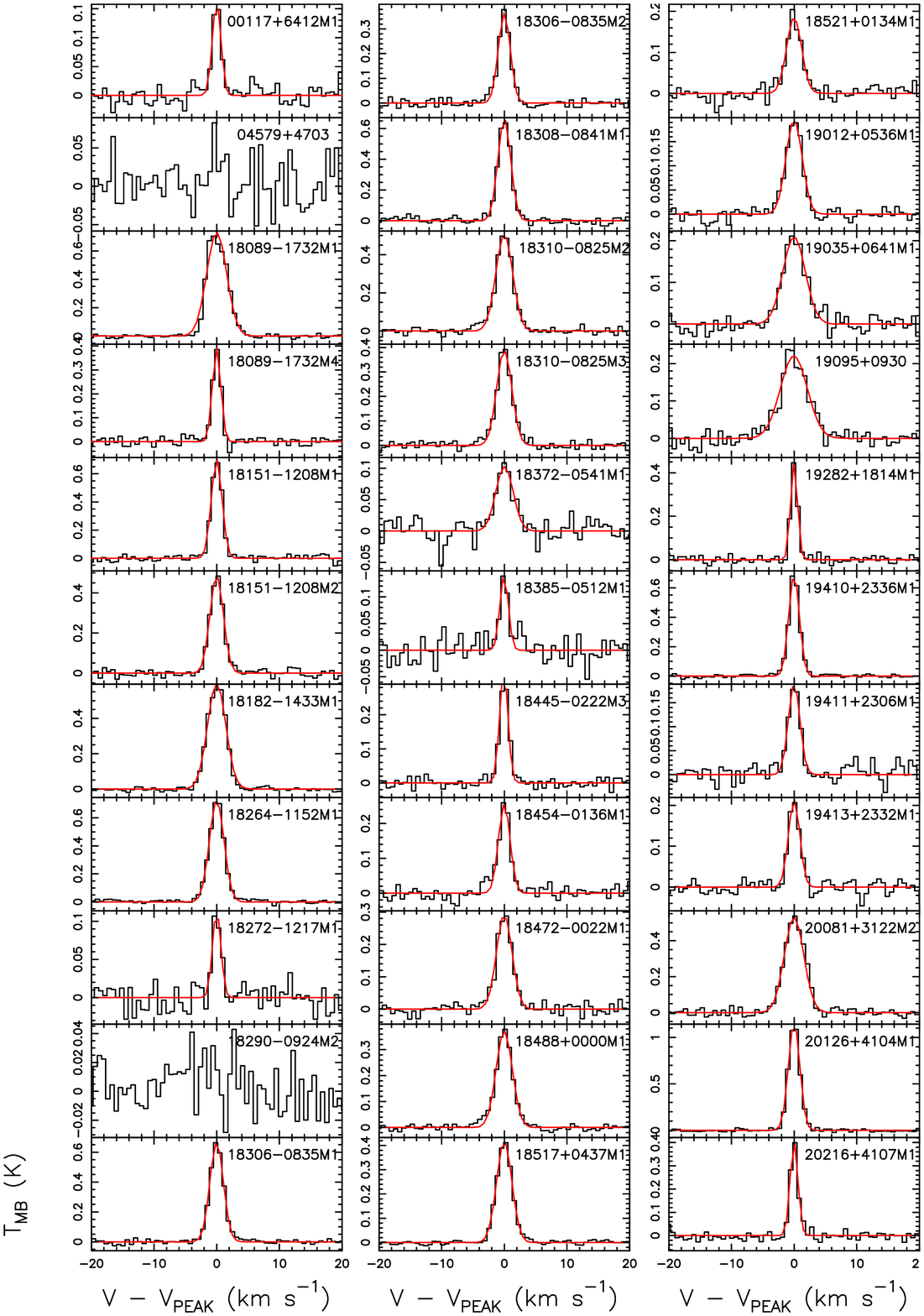}
\caption{ Spectra of HN$^{13}$C(1-0) obtained for the sources. For each spectrum the x-axis represents a velocity interval of $\pm20$ km s$^{-1}$ around the velocity listed in the second column of Table \ref{fit1}. The y-axis shows the intensity in main beam temperature units. The red curves are the best Gaussian fits obtained with MADCUBA.}
\centering
\label{hn13c-1}
\end{figure*}

\begin{figure*}
\centering
\includegraphics[width=30pc]{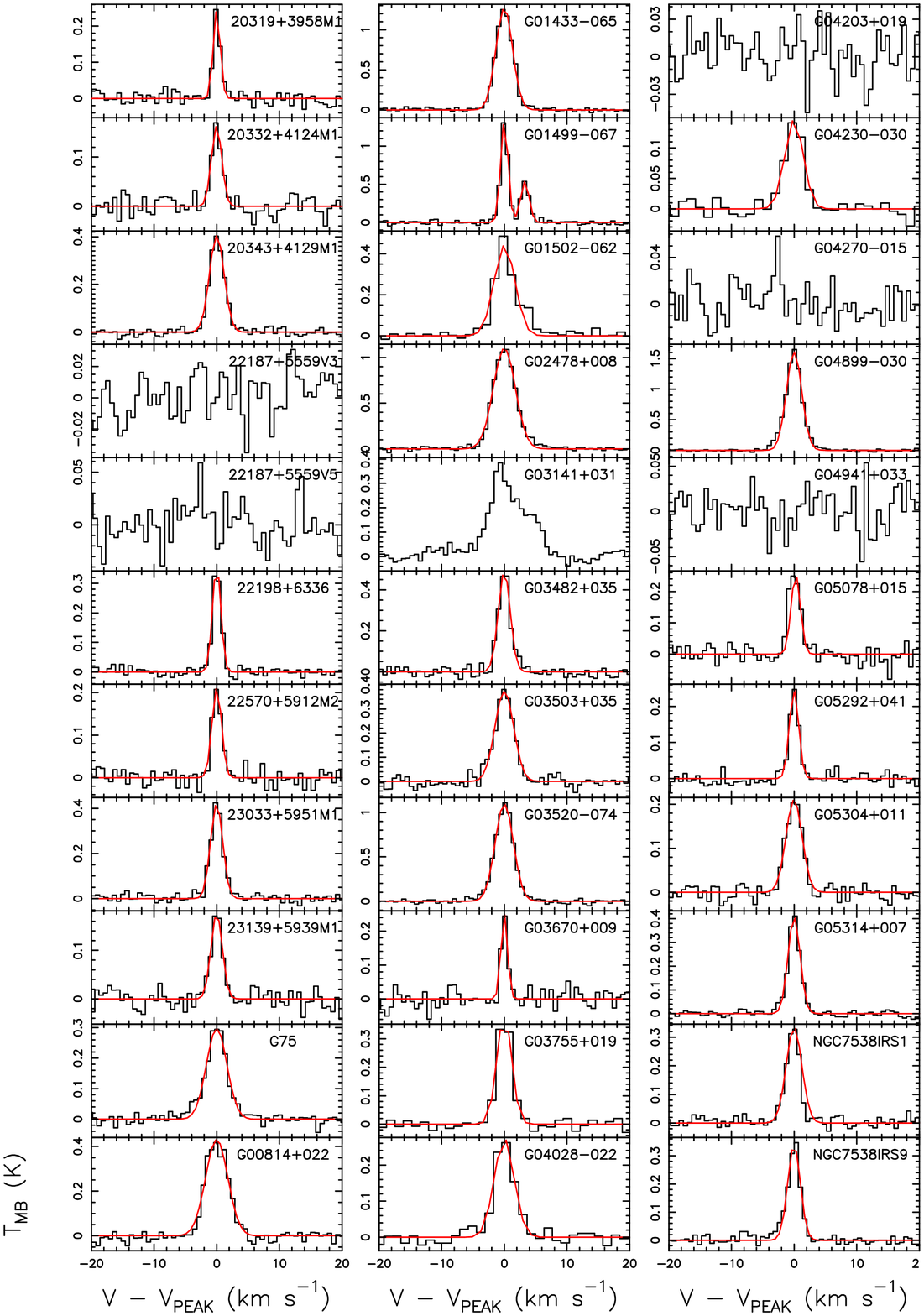}
\caption{Fig.~\ref{hn13c-1} continued. Note that for the source G01499-067 we have observed two components: we have fitted both lines and we have used both to compute the column densities.}
\centering
\label{hn13c-2}
\end{figure*}

\begin{figure*}
\centering
\includegraphics[width=30pc]{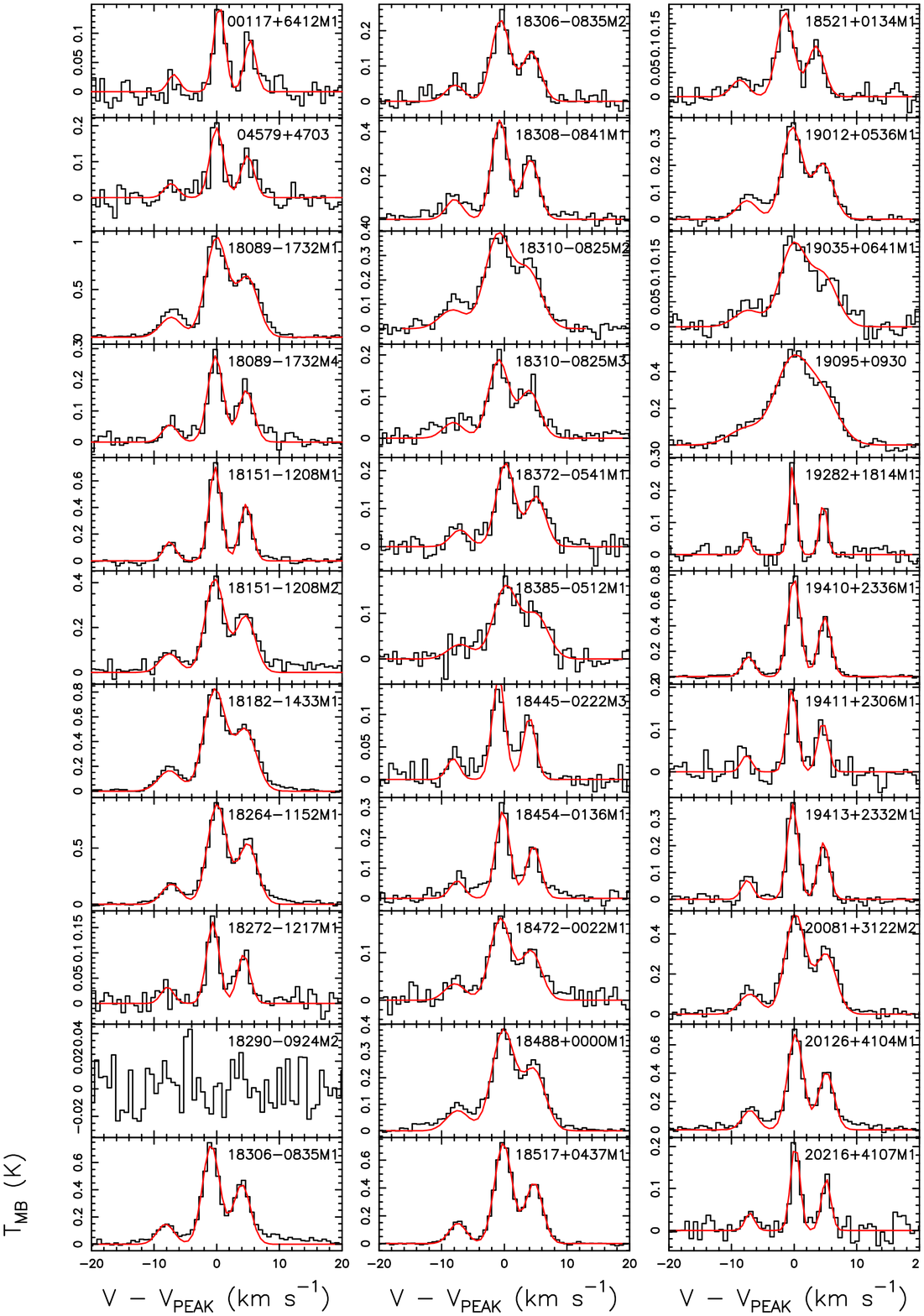}
\caption{Spectra of H$^{13}$CN(1-0) obtained for the sources. For each spectrum the x-axis represents a velocity interval of $\pm20$ km s$^{-1}$ around the velocity listed in the second column of Table \ref{fit2}. The y-axis shows the intensity in main beam temperature units. The red curves are the best Gaussian fits obtained with MADCUBA.}
\centering
\label{h13cn-1}
\end{figure*}

\begin{figure*}
\centering
\includegraphics[width=30pc]{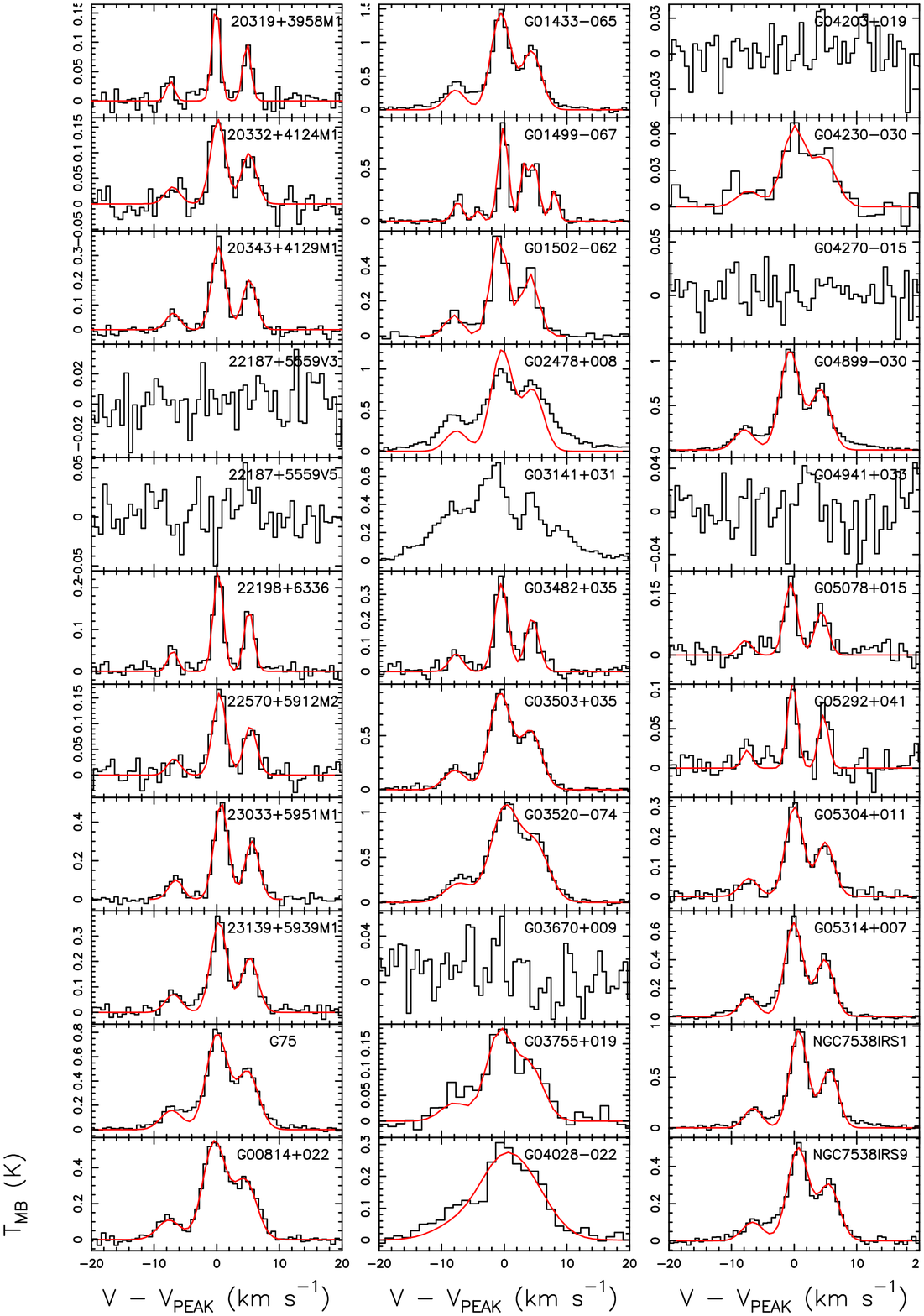}
\caption{Fig.~\ref{h13cn-1} continued. Note that for the source G01499-067 we have observed two components: we have fitted both lines and we have used both to compute the column densities.}
\centering
\label{h13cn-2}
\end{figure*}

\begin{figure*}
\centering
\includegraphics[width=30pc]{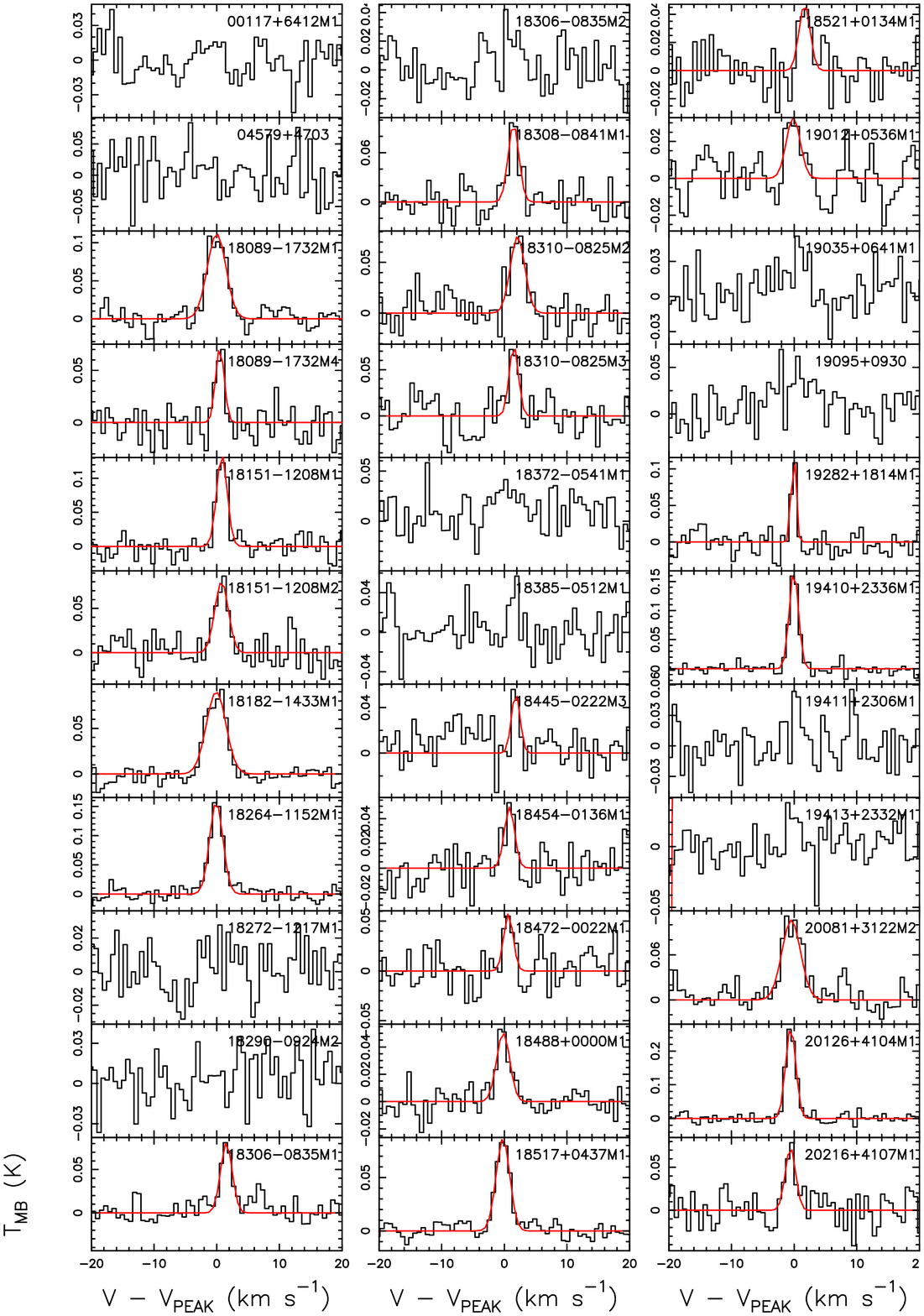}
\caption{ Spectra of H$^{15}$NC(1-0) obtained for the sources. For each spectrum the x-axis represents a velocity interval of $\pm20$ km s$^{-1}$ around the velocity listed in the second column of Table \ref{fit1}. The y-axis shows the intensity in main beam temperature units. The red curves are the best Gaussian fits obtained with MADCUBA.}
\centering
\label{h15nc-1}
\end{figure*}

\begin{figure*}
\centering
\includegraphics[width=30pc]{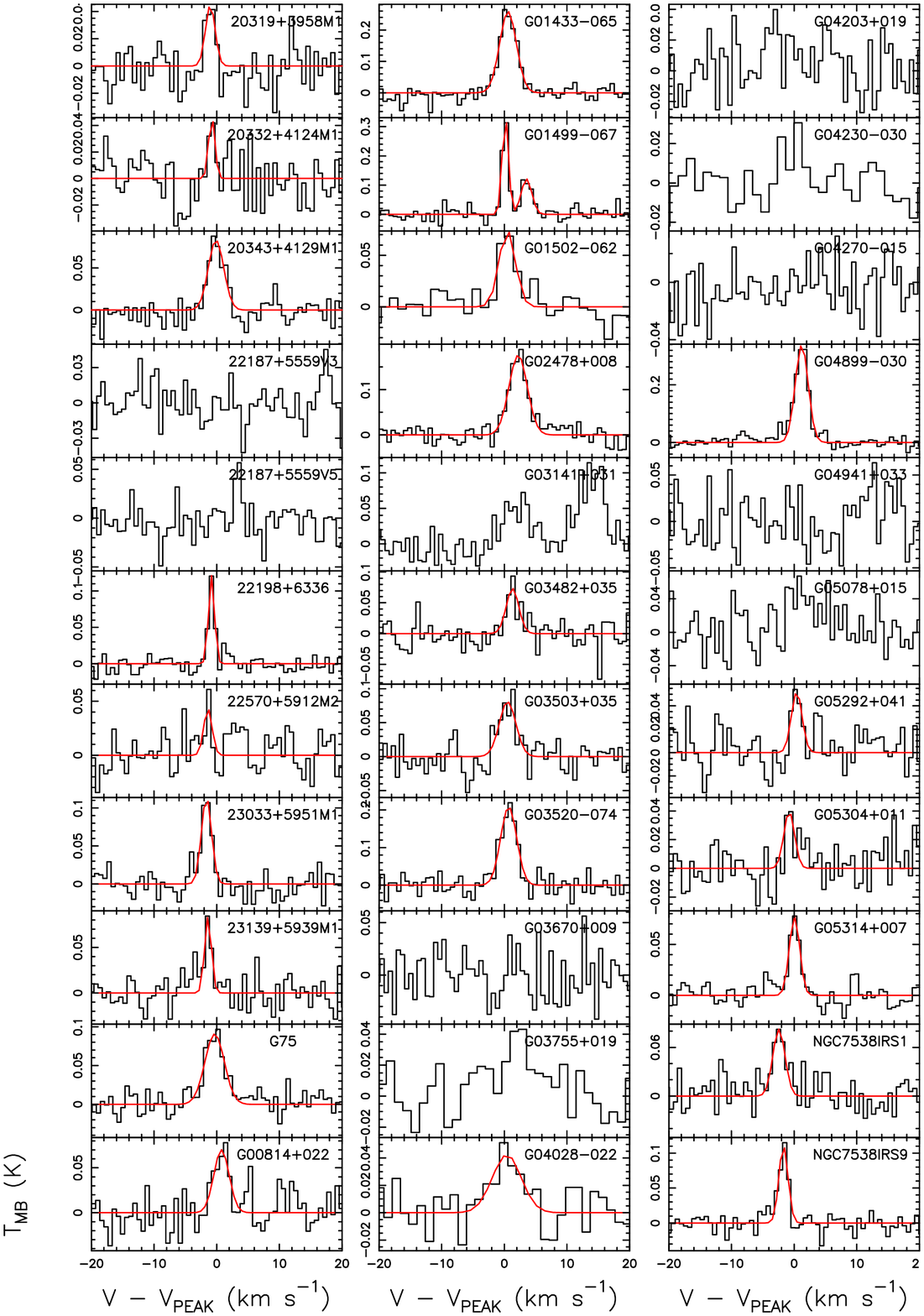}
\caption{Continue of Fig.~\ref{h15nc-1}. Note that for the source G01499-067 we have observed two components: we have fitted both lines and we have used both to compute the column densities.}
\centering
\label{h15nc-2}
\end{figure*}

\begin{figure*}
\centering
\includegraphics[width=30pc]{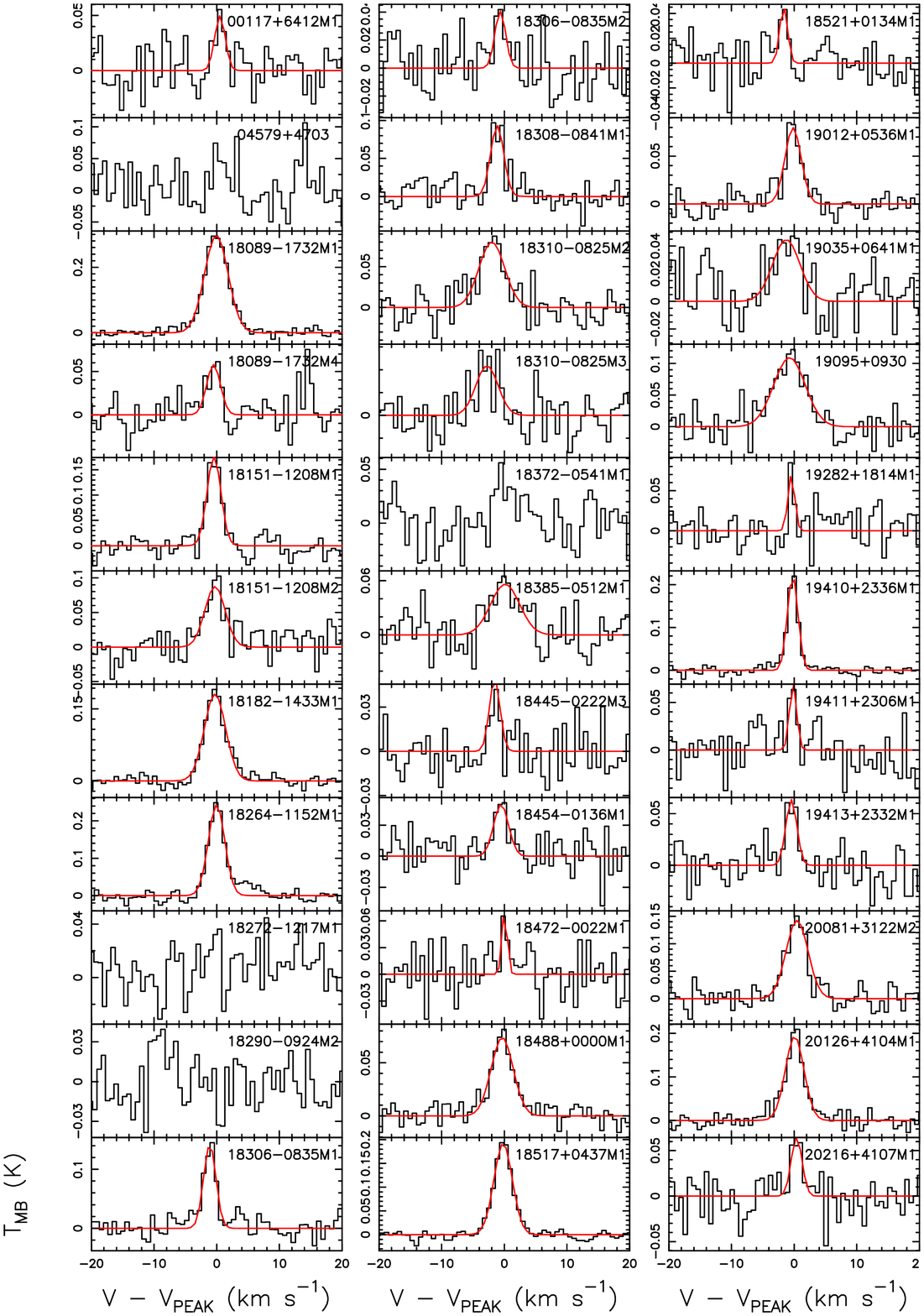}
\caption{ Spectra of HC$^{15}$N(1-0) obtained for the sources. For each spectrum the x-axis represents a velocity interval of $\pm20$ km s$^{-1}$ around the velocity listed in the second column of Table \ref{fit2}. The y-axis shows the intensity in main beam temperature units. The red curves are the best Gaussian fits obtained with MADCUBA.}
\centering
\label{hc15n-1}
\end{figure*}

\begin{figure*}
\centering
\includegraphics[width=30pc]{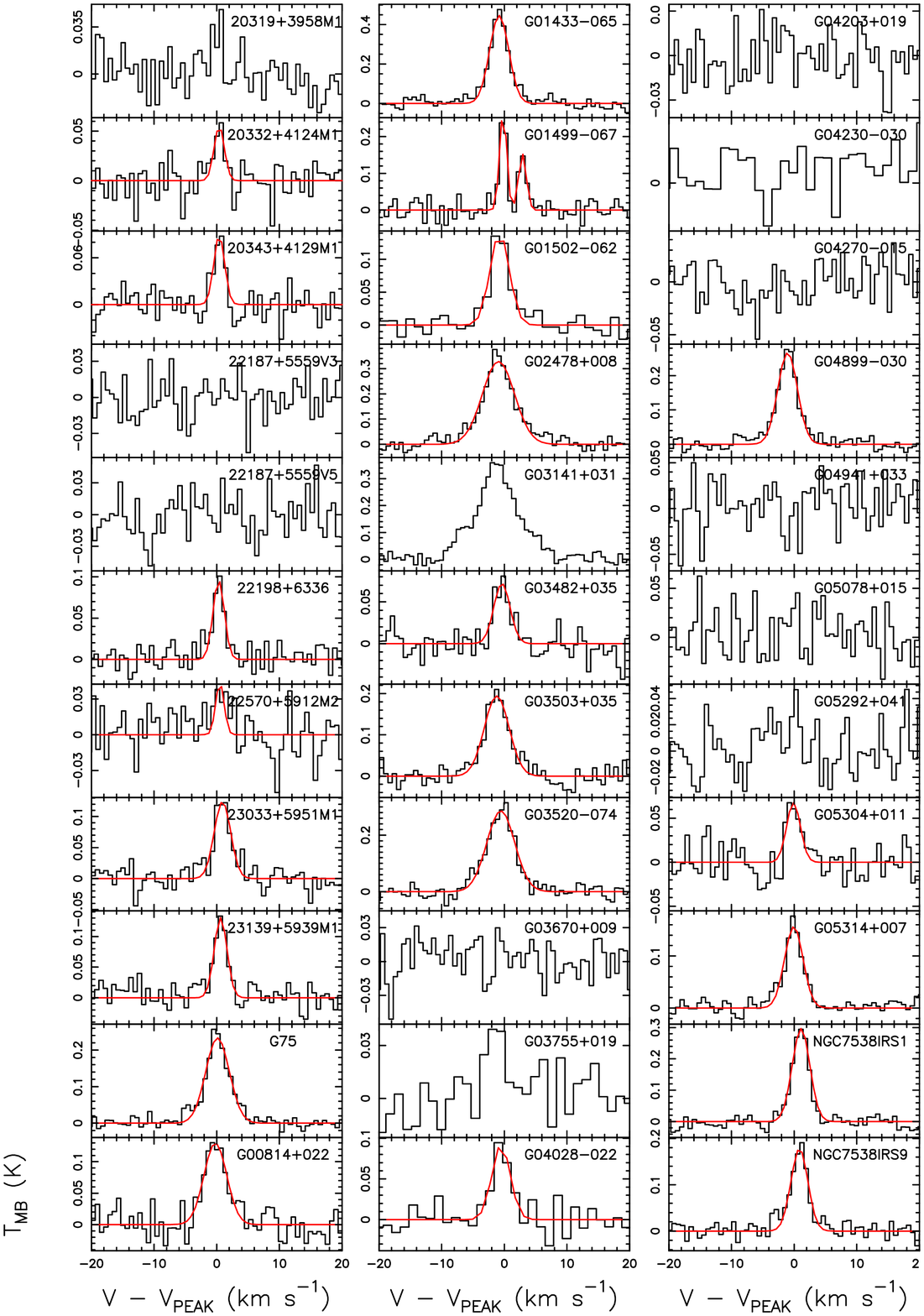}
\caption{Continue of Fig.~\ref{hc15n-1}. Note that for the source G01499-067 we have observed two components: we have fitted both lines and we have used both to compute the column densities.}
\centering
\label{hc15n-2}
\end{figure*}


\bsp	
\label{lastpage}
\end{document}